\documentclass[showpacs,amsmath,preprint,pre]{revtex4}
\usepackage{graphicx}

\begin{document}

\title{Adsorption hysteresis and capillary condensation  in
disordered porous solids: a density functional  study.}

\author{E. Kierlik\( ^{1} \), P. A.  Monson\( ^{2} \),
M. L. Rosinberg\( ^{1} \), and G. Tarjus\( ^{1} \)}
\affiliation{ $^{1}$Laboratoire de Physique Th{\'e}orique des
Liquides\cite{AAAuth}, Universit{\'e} Pierre et Marie Curie, 4 Place Jussieu,\\ 75252 Paris Cedex 05, France \\$ ^{2}$Department of Chemical Engineering, University of Massachussetts, Amherst,
MA 01003, USA}

\date{\today}

\begin{abstract}
We present a theoretical study of capillary condensation of fluids
adsorbed in mesoporous disordered media. Combining mean-field
density functional theory with a coarse-grained description in terms
of a lattice-gas model allows us to investigate both the
out-of-equilibrium (hysteresis) and the equilibrium behavior. We show
that the main features of capillary condensation in disordered solids
result from the appearance of a complex free-energy landscape with a
large  number of metastable states. We detail the numerical
procedures for finding these states, and the presence or absence of
transitions in the thermodynamic limit 
is determined by careful finite-size studies.
\end{abstract}
\maketitle

%\vspace{1cm}
%\hrulefill\
%\narrowtext

\def\be{\begin{equation}}
\def\ee{\end{equation}}
\def\bea{\begin{eqnarray}}
\def\eea{\end{eqnarray}}
\def\bit{\begin{itemize}}
\def\eit{\end{itemize}}

\section{Introduction}

Capillary condensation  of a gas in a single, infinitely long pore of
simple geometry (e.g., a slit) is a  genuine first-order phase 
transition that corresponds to the shift  of the bulk gas-liquid
transition due to confinement and adsorption on the solid walls
\cite{E1990}.
The jump  in the mean  density of the confined fluid occurs  at
a vapor pressure smaller than the bulk saturation value $P_{sat}$ 
and disappears above a capillary critical temperature 
$T_{cc}$ that is lower than the bulk critical temperature $T_c$.
As is usual with first-order phase transitions, the existence  of
local free-energy minima associated with the competing phases may result in 
 hysteretic behavior as the control variable (here,
the vapor pressure or, equivalently, the chemical potential) is swept upward and
downward. In a mean-field description 
of the phase transition, the metastable portions of the gas and
liquid branches exist only below $T_{cc}$, and the hysteresis loop is 
similar to that  in the van der Waals equation of state of a 
subcritical  bulk fluid.

This theoretical picture is, however, inadequate to describe the experimental situation of fluids  adsorbed in mesoporous solids
like porous glasses or silica gels\cite{GGRS1999}. 
Indeed, these amorphous materials contain  a
highly interconnected, irregular, three-dimensional pore network, and 
one  observes  at low temperatures a  steep increase of the adsorbed
quantity at pressures below $P_{sat}$, but no sharp vertical jump
that  would be the undisputable signature
of a  first-order transition (a noticeable exception is the case of 
 very dilute aerogels\cite{WC1990,Z1996}).  
There are also no density fluctuations  on large length scales that would signal the  onset of criticality. 
The main phenomenon is a hysteresis in the sorption isotherms between filling
and draining, with a marked asymmetry between the adsorption
and the desorption branches. The hysteresis loop shrinks in size as the
temperature increases and eventually disappears
above a certain temperature $T_h$  lower than $T_c$. There is also a whole
hierarchy  of subloops and scanning curves  that are obtained by
performing incomplete filling-draining cycles
\cite{B1963}.  Hysteresis loops are  perfectly reproducible on
the time-scale of usual  adsorption experiments  and are  widely
used for  characterization of porous materials \cite{GS1982}, despite
the fact that  a satisfactory understanding of the physical mechanism  is still missing. 
In particular, the question of whether  hysteresis results solely from 
metastability  in each  pore or from the
interconnectivity of the pore network is
still a matter of debate\cite{BE1989}. Also, the connection between hysteresis (which is intrinsically an
out-of-equilibrium phenomenon) and a possible 
underlying equilibrium transition remains largely unexplored. In fact,
the absence of a sharp jump in the adsorption isotherm is usually 
taken as the indication that there is no thermodynamic phase transition
in the porous material, the continuous filling being  a
consequence of the distribution of pore sizes and
shapes. This can be justified  in
the framework of an independent-pore model\cite{BE1989,E1967,LH2001} in which each  pore is regarded as an isolated
system, much like in the Preisach model of
hysteresis in magnetic systems\cite{P1935}. The global sorption isotherms are then obtained by performing an
average over the distribution of pore sizes, an operation that
smears out any  discontinuous behavior of the elementary
units. The independent-pore model, however, is a phenomenological
description, with no microscopic justification, and it is in general
impossible  to characterize geometrically the independent domains.

In order to better understand these important issues,  we have
recently undertaken a detailed investigation of the equilibrium and
out-of-equilibrium (hysteretic) properties of a  lattice model
 introduced some years ago as a coarse-grained description of
fluids in disordered solids\cite{PRST1995}. Although simple, the model contains the essential physical ingredients that characterize
such  systems  (confinement, randomness of the pore
network, wettability of the solid surface) and it has a well-defined
Hamiltonian, which  allows to perform a statistical-mechanical
calculation of  the equilibrium properties.  This contrasts with previous phenomenological
descriptions  of  capillary condensation that use predefined kinetic rules for
the evolution of the fluid configurations
in the pore space\cite{M1982}. Like in most theoretical studies of capillary
condensation  (see, e.g., Ref.\cite{E1990} and references therein),
our analysis is based upon mean-field density functional theory (i.e.,
local mean-field theory in the terminology of  lattice models). As
shown in preliminary reports on this work\cite{KRTV2001,KMRST2001}, 
our theoretical predictions qualitatively reproduce
many aspects of the phenomenology  of capillary
condensation, in particular  the asymmetry of the hysteresis loops, their
temperature dependence, and the shape of the scanning curves. This
demonstrates the relevance of the model and supports the hypothesis
that thermal fluctuations, which are ignored  in a mean-field
treatment, play a minor role on the time scale of adsorption
experiments (a feature confirmed by Monte-Carlo 
simulations performed on related models\cite{SM2001,WM2002}).  Our results suggest that  the
experimental observations mainly reflects the complexity of the
underlying free-energy
landscape at low temperatures. Indeed, quite similarly to disordered
magnetic systems, such as spin-glasses or
random-field spin models, that 
have been extensively studied in recent years\cite{N1997}, we find a large number of metastable
states below a certain temperature. As  one varies  the external control parameter 
(here, the chemical potential), the system
either follows a given local minimum of the free-energy surface or,
when the local minimum looses its stability, jumps to another, nearby minimum. This
results in a hysteretic behavior. It also causes a 
loss of thermodynamic consistency along the sorption
isotherms, a feature that  cannot be explained in the framework of the classical van der
Waals picture of metastability where only two free-energy minima are
present below
$ T_c$. As far as the equilibrium properties are concerned, we find  that a genuine thermodynamic 
phase transition may occur  when the perturbation induced by the solid
is sufficiently small, which is at odds with the picture built upon
the independent-pore model.  Moreover, the existence of this underlying
transition cannot be deduced from  the behavior of the
adsorption branch that may be either continous or
discontinuous. 

The goal of the present work is to give a general picture of capillary
condensation in disordered mesoporous materials, to detail some of the
mean-field density functional theory (DFT) calculations,
and to report on new results. The paper is arranged  as follows. In Section II we review the model
and  the theory, and we describe the numerical procedures  used
to find the solutions of the DFT equations. 
Section III presents the results for the 
hysteresis loops and the scanning curves. Section IV 
is devoted to a detailed analysis of the DFT equilibrium
isotherms;  a careful finite-size scaling study is performed to
check on the presence or absence of a sharp transition in the infinite
volume limit. We summarize our main findings in  Section V.

\section{Model and theory}

\subsection{Model}

We consider  a three-dimensional lattice where each of the  $N$  sites
($i=1,2,..N$) may be occupied by a fluid or a matrix
particle,  as described by the occupancy variables $\tau_{i}$ and
$1-\eta_{i}$, respectively ($\tau_{i}=1$  if  site $i$ is   occupied by a
fluid particle and $\tau_{i}=0$ if it is not, $\eta_i=0$ if site $i$ is  occupied by a matrix
particle and $\eta_i=1$ if it is not). Multiple occupancy of a site is
forbidden and only nearest neighbor  (n.n.)
interactions are taken into account. The associated Hamiltonian is\cite{PRST1995,KRTP1998}

\begin{equation}
{\cal H} = -w_{ff}\sum_{<ij>} \tau_{i}\tau_{j} \eta_i \eta_j -w_{mf}\sum_{<ij>} [\tau_{i}\eta_i (1-\eta_j)+\tau_{j}\eta_j (1-\eta_i)] 
\end{equation}	
where  $w_{ff}>0$ and $w_{mf}$ denote the fluid-fluid and matrix-fluid
interactions, respectively, and the sums run over distinct  n.n. pairs. 
Fluid particles are in thermal
equilibrium with an external  reservoir that fixes their chemical potential
$\mu$ and the temperature  $T$, whereas matrix particles are distributed
according to some given probability distribution function $P(\{\eta_i\})$.  In the simplest
version of the model that is considered in this work, the matrix particles
are distributed randomly on the lattice with the canonical 
constraint that  $\sum_i\eta_i=(1-\rho_m)N$, where $\rho_m$ is the matrix density.

This is  clearly a coarse-grained description  that leaves aside most
of  the microscopic details of an actual solid-fluid system. However, 
experiments show that fluids in  disordered media
share some generic qualitative properties. These latter can then be captured by
 a simple model, with the great advantage that the model is amenable to detailed
numerical investigations. The model has only two parameters that can be
tuned independently: $1-\rho_m$, the porosity of the matrix
 (here, dilution plays the role of confinement), 
and $y=w_{mf}/w_{ff}$, the interaction  ratio that  controls the wetting properties of the 
solid-fluid interface. One can easily improve the
description, for instance by using a distribution of the 
matrix particles that is more faithfull to the microstructure of a real solid\cite{SM2001,WSM2001,DKRT2002}. 

By transforming the above lattice-gas Hamiltonian to its
equivalent spin-$\frac{1}{2}$ Ising form\cite{KRTP1998}, one finds
 that the value $y=1/2$ plays a special role. Indeed,  at
fixed  $\rho_m$, there is a  symmetry $y \leftrightarrow1-y$ expressed by
\be
\rho_i(y,\mu,T)+\rho_i(1-y,-\mu-cw_{ff},T)=\eta_i \,
\ee
where $c$ is the coordination number of the lattice and $\rho_i=<\tau_i\eta_i>$ is the average fluid density at site $i$ for 
a given realisation of the random matrix defined by the set $\{\eta_i\}$. 
(Throughout the paper we denote by $<..>$  the thermal average for a
given matrix sample and by $[..]$ the ensemble average over
the different matrix samples.) This property (that is special to the
lattice-gas description) 
allows  us to restrict the study to $y\geq 1/2$ and to attractive solid-fluid
interactions only. 
It is important to note  that $y=1/2$ is the only case where the
hole-particle  symmetry of the bulk lattice gas  is
preserved. The Hamiltonian described by Eq. (1) is then equivalent to that
of the site-diluted Ising model, a model for disordered
magnets whose properties are  well-documented\cite{S1983}. In
particular, this mapping tells us immediately that the system undergoes a liquid-vapor phase separation at low
temperature  when the porosity  is  large enough.
Specifically, the critical temperature $T_{cc}(\rho_m,y=1/2)$ decreases monotonically
from  $T_c$ to $0$ as $\rho_m$  varies from $0$ up to $1-p_c$, where $p_c$ is the
site-percolation threshold of the lattice. Moreover, because of  the 
hole-particle
symmetry, the transition takes place at the same  chemical potential as the 
bulk liquid-gas transition, $\mu=\mu_{sat}=-cw_{ff}/2$.

The physical behavior for $y\neq 1/2$ is more  complicated, but more
relevant to real adsorption experiments; then, random fields come into play in addition to
dilution, which breaks  the up-down (or hole-particle) symmetry of the
model\cite{KRTP1998,SP1992}. These random  fields are  spatially
correlated (at a local scale, though) and can take $c+1$
distinct  values at a fluid site $i$, depending on the number $n_i$ of n.n sites that
are occupied by a matrix particle ($0\leq n_i\leq c$). For a purely random
matrix, one has $P(n_i)=\binom{c}{n_i}\rho_m^{n_i}(1-\rho_m)^{c-n_i}$ which
leads to a binomial distribution of the fields\cite{MSCCCB1991}. 
On general grounds, one expects that the presence of random fields has  dramatic consequences
for the out-of-equilibrium properties of the model (see, e.g.,
Ref.\cite{FGK1988}). According to  the
original argument of  Imry and Ma\cite{IM1975} based on the energy 
balance for domain formation, this sort of
randomness should not forbid the system from undergoing a
sharp thermodynamic transition  at low temperature in three
dimensions; but it should strongly alter the critical properties\cite{N1997}.  Specifically,  as will be illustrated by the
calculations of  section IV,  we expect that there is a nonzero critical
temperature  $T_{cc}(\rho_m,y)$ 
when  the  effective strength of the disorder is weak, i.e.,   when $\rho_m$ is
smaller than a certain value $\rho_m^{max}(y)$ (or $y$ smaller than
$y_{max}(\rho_m))$. Moreover, since the random fields 
have a  mean that is strictly positive for $y>1/2$, the jump in the
fluid density for
$T<T_{cc}(\rho_m,y)$ occurs at a chemical potentiel that is {\it lower} than $\mu_{sat}$: this first-order phase
transition is thus  a genuine equilibrium capillary
condensation. This leads to the putative phase diagram  shown
in Fig. 1  where the boundary between  the two
regions  corresponds to $T_{cc}(\rho_m,y_{max}(\rho_m))=0$. (In fact, as will be discussed below, the  behavior of
{\it finite} matrix samples is rather complicated and one cannot discard the possible
existence of  two or more phase transitions occuring at 
different chemical potentials, as found in a previous
work\cite{KRTP1998}). 

\subsection{Mean-field density functional theory}

There are  essentially two different approaches  for studying the statistical
properties of fluids or magnets in the presence of quenched
disorder. The first one is the replica method in which only
quantities that are averaged over the disorder can be calculated. This
is the method that we have used 
in our preceding studies of fluids in porous
solids\cite{PRST1995,KRTP1998,KMRT1997} where we have derived formal equations for 
the fluid-fluid and matrix-fluid average (hence, translationally invariant) pair correlation functions.
In principle, the solution of these equations, supplemented by some 
appropriate closure approximations, yields
the equilibrium properties of the model in the thermodynamic limit. 
However, standard  approximations of liquid-state theory  may run into difficulties in the
presence of a large number of metastable states (as is the case here).
Moreover, the physical content of this
formulation is not  very transparent, especially if replica symmetry breaking occurs.
The second method consists in first calculating the properties of  a single finite
sample and averaging the results over the disorder at a later stage
of the calculation, as was done by Thouless-Anderson-Palmer in their
study of the infinite-range Ising spin-glass model\cite{TAP1977}. This is the
approach  that we choose in the present  work because  it  
allows us to investigate the free-energy landscape for a given disorder
realization and to understand its relation to the 
hysteretic behavior of the model. The main inconvenience of working with finite
samples is that  a careful finite-size scaling study is needed in
order to conclude on  the existence of sharp phase transitions in the
infinite-volume limit. This usually  requires  a significant amount of numerical analysis. 

We thus consider a finite lattice of linear size $L$ and, in order to simplify
the numerical work and to minimize surface effects, we use
periodic boundary conditions  in all directions. The adsorbed fluid is then
statistically homogeneous, with a mean
density $\rho_f=(1/N)\sum_i  \rho_i=[<\tau_i\eta_i>]$ in the limit of a large system.
As discussed  elsewhere\cite{SM2002,KRT2002}, choosing periodic boundary conditions
is not completely benign: it may alter dramatically the nature of
the desorption process, but has no consequences for the equilibrium
properties nor, for the problem at hand, for the adsorption process.
 
For a given realization of the random matrix, the DFT starts with
the expression of the grand-potential functional of the fluid-density
field (on a lattice, the
functional is actually a function of the $\{\rho_i\}$'s)\cite{KRTV2001,KMRST2001}:
\bea
\Omega(\{\rho_i\})&=&k_BT \sum_i[\rho_i\ln \rho_i+(\eta_i-\rho_i)\ln(\eta_i-\rho_i)]\nonumber\\
&-&w_{ff} \sum_{<ij>}\rho_i\rho_j -w_{mf}\sum_{<ij>}[\rho_i(1-\eta_j)+\rho_j(1-\eta_i)]-\mu\sum_i\rho_i \ ,
\eea
an expression that is obtained as usual in a mean-field approximation
by neglecting correlations between the thermal fluctuations of the
instantaneous fluid densities, $\tau_i\eta_i-<\tau_i\eta_i>$. It should be
stressed, however, that the present approach fully accounts for the
disorder-induced fluctuations, fluctuations that are expected to be
the dominant ones in systems with random fields\cite{N1997}. It also
preserves all geometric constraints and, as a result, properly
describes the site percolation threshold. 

Minimization with respect to the $\{\rho_i\}$'s leads to the following equations:
\begin{subequations}
\be
\rho_i=\eta_i[1+e^{-\beta v^{eff}_i}]^{-1}
\ee
where $\beta=1/(k_BT)$ and $v^{eff}_i$ is the effective  potential at site $i$,
\be
v^{eff}_i=\mu+w_{ff}\sum_{j/i}[\rho_j+y(1-\eta_j)] \ ,
\ee
\end{subequations}
where the sum runs over the $c$ nearest neighbors of site $i$. It is
easy to see  that these $N$ coupled non-linear 
equations are invariant in the change $\mu \leftrightarrow-\mu-cw_{ff},y \leftrightarrow1-y, \rho_i
\leftrightarrow\eta_i-\rho_i$, which shows that the mean-field DFT is faithfull to the symmetry property
expressed by Eq. (2). There are in general several solutions to
Eqs. (4), that can be maxima, saddle-points, as well as minima of the
grand-potential functional. In what follows, we focus only on the
metastable states, i. e., on the minima. We label these latter by the superscript $\alpha$. 
The grand potential for the solution $\{\rho_i^{\alpha}\}$ is then given by\cite{KRTV2001}

\be
\Omega^{\alpha}=k_BT \sum_i \eta_i\ln(1-\frac{\rho_i^{\alpha}}{\eta_i})+w_{ff} \sum_{<ij>}\rho_i^{\alpha}\rho_j^{\alpha}  \ .
\ee

The  adsorbed  fluid is in equilibrium  with a bulk fluid whose uniform
density $\rho_f^{bulk}$ satisfies the standard  mean field equation of
state of a n.n. lattice gas, which yields a  critical  point  located  at  $k_BT_c/w_{ff}=c/4$ and
$\rho_f^c=1/2$.

For a given matrix realisation and a given temperature, 
Eqs. (4) were solved by the simplest iterative method: an initial
set of local densities, $\{\rho_i^{(0)}\}$, was used to calculate a set of effective
local potentials, $\{v^{eff(0)}_i\}$,  from which a new set of
densities was generated, and so on, until a fixed point of the
iterative procedure was obtained. The main interest of such an
algorithm is that it automatically discards solutions of Eqs. (4) that
are not minima of the grand potential\cite{LBL1983}.

Two types of numerical
calculations were performed. First, to mimic the protocol of sorption 
experiments, we progressively increased (respectively, decreased) the chemical potential from
a large negative (respectively, from the bulk saturation) value, using at each subsequent $\mu$
the converged values of the
$\rho_i$'s at the previous chemical potential to start the iteration  (the
elementary step was $\Delta \mu/w_{ff}=10^{-2}$ or $10^{-3}$). At
each $\mu$, convergence was assumed when $(1/N)
\sum_i(\rho_i^{(n+1)}-\rho_i^{(n)})^2<10^{-8}$, where the superscript $(n)$
denotes the nth iteration. More complicated chemical
potential histories were also considered to describe the
adsorption and desorption scanning curves or the inner hysteresis
loops. Secondly, to search for other solutions of Eqs. (4), we
generated at  each  $\mu$ a certain number of
initial configurations (typically, $10^{2}$) corresponding to {\it uniform} fillings of the lattice with
different overall fluid densities
(i.e. $\rho_i^{(0)}=\rho_f^{0}$). Our objective was not an exhaustive enumeration of all solutions. Indeed, a more systematic search would require  the use of
non-uniform seed configurations such as random or chequerboard
configurations\cite{LMP1995,MPP2000}. However, our set of
initial configurations was in general sufficient
to obtain a good approximation of the equilibrium solution
(i.e., that giving  the lowest value of $\Omega$) in the 
range of temperatures explored in the present work. This search required 
a rather stringent convergence criterion: the iteration
algorithm was stopped when $(1/N)\sum_i(\rho_i^{(n+1)}-\rho_i^{(n)})^2<10^{-14}$, and two solutions
$\{\rho_i^{\alpha}\}$ and $\{\rho_i^{\beta}\}$ were considered as different
whenever $\sum_i(\rho_i^{\alpha}-\rho_i^{\beta})^2>10^{-6}$. (We indeed
observed, as in Ref.\cite{LMP1995}, that different initial conditions
could yield the same final configuration.) 

The above mean-field DFT (or local mean-field) equations have an interesting property that is
worth pointing out. 
Suppose that for a given realization $\{\eta_i\}$,
two different sets $\alpha$ and $\beta$ of  fluid densities satisfy the condition
$\rho_i^{\alpha} \leq \rho_i^{\beta}$ for each site in the system (this is
of course a very special ordering: most fluid configurations do not
have such a relationship). Then, because of the convexity of the
exponential function, the densities on the left-hand side of Eq. (4a) satisfy
the same ordering. This ordering is thus automatically preserved by
the  iteration algorithm. In particular, any
solution of Eqs. (4) yields a mean fluid density $\rho_f$ that is larger
(respectively, lower) than the density obtained by starting the iteration
procedure from an initially empty (respectively, filled) lattice. This defines two
extremal curves that coincide with the adsorption and desorption isotherms.
For the same reason, the algorithm satisfies a ``no-passing'' rule\cite{M1992}
which implies a so-called return-point memory property: when the chemical
potential is adiabatically (i.e., very slowly) swept upward,
downward and then upward again (or conversely) so as to
come back to its original value, the system returns to exactly the same 
configuration of the $\rho_i$'s (and thus to the same mean
density $\rho_f$). This is fully equivalent to the return-point memory property
observed in ferromagnetic systems, and the demonstration of the
``no-passing'' rule is similar to that given by Sethna {\it et
al.}\cite{S1993} for the athermal dynamical response of the
random-field Ising model to an external field. 
Clearly, using an iterative numerical mean-field scheme is analogous
to having  a  zero-temperature dynamics:
the system only evolves  under the influence of the external control
parameter (magnetic field or chemical potential), and there is no possible equilibration mechanism via thermally activated
processes that would take the system from one metastable state to
another. In the absence of a satisfactory theoretical treatment of
these effects, it  is the comparison to computer simulations
or to experiments  that can tell us whether  the qualitative picture emerging from
this calculation is modified or not by thermal fluctuations\cite{WM2002}.

For various system sizes, Eqs. (4) were solved for many matrix
samples, and the results  were then  averaged over this set of matrix realizations. The  number of samples
depended on the lattice size and on the property studied.
To obtain the hysteresis loops and the  scanning
curves, it was sufficient to use a small number of realizations.
 Indeed, the determination  of the adsorption/desorption
isotherms did not require a large amount of computational effort  so
that rather  large samples (with typically  $L=48$) could be used\cite{rem}.
 On the other hand,
the search for the equilibrium isotherms was  much more demanding,
and  we had to consider smaller systems ($L=8,10,12,$ and $16$) to
perform the finite-size study described in Sec. IV. The model has a
good self-averaging behavior far from criticality and 
good statistics could be reached with a few hundreds samples.

All the results presented here are obtained for  a bcc
lattice ($N=2L^3, c=8$). The present model has a rich behavior and it would be
interesting to perform  a systematic study of its properties as a function of $\rho_m,y$ and
$T$. This task, unfortunately, would require a considerable amount of computational
work. We are thus limited to select a few points in the parameter
space, which we think  represent typical situations. We consider
 a single value of the matrix density, $\rho_m=0.25$, which is just above the site percolation threshold
of  the lattice, $p_c=0.243$\cite{SA1994}, and a single temperature, $T^*=k_BT/w_{ff}=0.8$, which is well
below the mean-field critical temperature of the bulk fluid,
$T_c^*=2$. We defer the 
 study of the influence of the temperature and the porosity on the 
behavior of the adsorbed fluid to a future work in which the matrix sites
will be distributed over the lattice in a non-random and more
realistic fashion\cite{DKRT2002}.

\section{Hysteresis loops and scanning curves}

We first consider the hysteretic, out-of-equilibrium behavior of the
system which can be described within the mean-field DFT by using the
first protocol described above. Representative adsorption and
desorption isotherms are shown in Fig. 2 where the mean fluid density
$\rho_f$ is plotted versus the reduced chemical potential  $\mu^*=\mu/w_{ff}$
for several values of the interaction ratio $y$. When $y$ increases,
i. e., when the matrix-fluid interaction gets stronger, more fluid can
be reversibly adsorbed in the vicinity of the matrix particles, and
the hysteresis loop shrinks and occurs further away from the bulk
saturation value, $\mu^*=-4$.

Scanning curves obtained by performing incomplete filling of the
matrix and then decreasing the chemical potential to drain the
adsorbed fluid (desorption scanning curves) or by the reverse
procedure of incomplete draining before filling (adsorption scanning
curves) are illustrated in Figs. 3a and 3b for $y=1.5$. As seen by
comparing the two figures, the shape of the curves
is markedly different on desorption and adsorption. The
theoretical predictions are strikingly similar to those measured  by
Brown\cite{B1963} for Xe in Vycor, a porous glass (see, e.g., Fig. 8 in
Ref.\cite{BE1989}), which supports our claim that the present
coarse-grained model contains the important physical ingredients to
describe adsorption in disordered porous media.
As stressed by Ball and Evans\cite{BE1989}, the shape of the predicted
scanning curves is indeed a rather stringent test for the validity of
any proposed model. In particular, the independent-pore model (or
Preisach model\cite{P1935} in the context of magnetic systems) does
not properly account for the observed curves. 

Connectivity of the pore network shows up
distinctly when considering hysteresis subloops obtained by performing
more complicated cycles of the chemical potential than those
considered so far. This is illustrated in Fig. 4. The two subloops
shown in the figure display two characteristic features: first, the
property of return-point memory already discussed and also satisfied
by the independent-pore model; and second, the lack of congruence,
which means that the two subloops, even though they open and close at
the same chemical potentials, cannot be superimposed on top of each
other by a mere translation along the vertical axis. Since,
irrespective of its detailed implementation, the independent-pore
model predicts exact congruence, a lack of congruence is a clear-cut
signature of the connectivity of the pore space\cite{WH2000}.

The physical reason for the presence of hysteresis loops, scanning
curves, and various subloops is the existence of many metastable
states in which the system gets trapped, at least on the time scale of
the experiment. If thermally activated processes that could allow for
untrapping from these metastable states are characterized by time
scales longer than the experimental time scale, the system can only
evolve under the influence of an external driving force, i.e., a
change in the chemical potential. This is precisely the situation
described by the mean-field DFT approach. The grand-potential
free-energy hypersurface (or, to use a more pictorial term,
``landscape'') defined by Eq. (3), $\Omega$  as a function of the $\{\rho_i\}$'s,
evolves with $\mu$: a given minimum can be continuously deformed and at some
point looses its stability and disappears. Hysteresis loops, scanning
curves, and other hysteresis subloops (as illustrated in Fig. 4)
correspond to various paths among minima of the grand-potential
landscape. To make this more visual, we have also plotted in Figs. 3a
and b the minima obtained from the DFT by following the second
protocal described in section II B.

The above landscape picture allows one to understand another puzzling
feature associated with hysteresis. Based on the standard van der
Waals description of metastability, it is often assumed that
thermodynamics can still be used to describe the behavior of the fluid
along the adsorption and desorption isotherms, even in the region of
hysteresis. However, as illustrated in Figs. 5a and b, again for
$y=1.5$, this assumption may fail in disordered porous materials. We
compare in these figures the mean fluid density $\rho_f$ directly
obtained from the solution of Eqs. (4) with that obtained from the
Gibbs adsorption relation, $\rho_f=-\left.\partial (\Omega/N) / \partial \mu\right|_T$, in which the
fluid grand potential is computed from Eq. (5) and the derivative is numerically calculated with a step
$\Delta \mu^*=0.01$. There is a clear violation of the thermodynamic
consitency along both the adsorption and the desorption branches of
the hysteresis. This can be rationalized by recalling that the system
jumps occasionnally from one local minimum of the grand potential to
another. Correspondingly, there are jumps in $\rho_f$ and $\Omega$. These
jumps may be extremely small (as indicated by the smooth curves
obtained for the adsorption and desorption isotherms), but the system then
looses  thermodynamic consistency between $\rho_f$ and
the change in $\Omega$. This point will be further discussed in the next
section on equilibrium properties.

All the above results were obtained for $y=1.5$. Qualitatively similar
behavior is obtained for other values of the interaction ratio, e.g.,
for $y=1$. Figs. 6a and b display the hysteresis loop, several
desorption and adsorption scanning curves, as well as metastable
states obtained by solving Eqs. (4) according to the second protocol
of section II.B. The main difference with the curves shown in Figs. 3a
and b for $y=1.5$ is the presence of a line of metastable liquid
states on the desorption branch that considerably widens the
hysteresis loop. This line, which is accompanied by no other
metastable states in a whole range of chemical potentials except the
gas-like states on the adsorption branch, is actually an artefact of
the periodic boundary conditions used in the present calculation. As
shown elsewhere\cite{SM2002,KRT2002}, these liquid-like states become unstable
as soon as one introduces a physical interface between the matrix and
the external reservoir.

\section{Equilibrium isotherms}

As pointed out in the preliminary reports on this work\cite{KRTV2001,KMRST2001}, the existence of
a complex free energy landscape  changes dramatically the 
description of  capillary condensation built  upon the
independent pore model and the standard  van der Waals picture of
phase transitions. In particular, determining the equilibrium
state that yields the lowest value ot the grand potential
becomes a non-trivial task, as explained in this section. For
simplicity, we again restrict our study to  the cases $y=1.5$ and $y=1$, two
values of the interaction ratio that  illustrate the system behavior
in the two regions of Fig. 1 corresponding  to the strong- and
low-disorder regimes, respectively. 

\subsection{The strong-disorder regime}

To  each solution $\{\rho_i^{\alpha}\}$ of the mean-field equations, obtained by the search procedure explained in section
II.B, corresponds a local minimum $\Omega^{\alpha}$ of the grand potential given by
Eq. (5). The result, of course, is sample-dependent. Fig. 7 shows the
reduced grand-potential densities $\omega^*=\Omega/(Nw_{ff})$ corresponding to
the metastable solutions  found for $y=1.5$ in a matrix of linear size $L=48$ (this is the same sample as in
Fig. 3). Two comments are in order. First, the curves for the adsorption and
desorption branches, $ \Omega^{ads}(\mu)$ and $\Omega^{des}(\mu)$, cross each other
at some value of the chemical potential,  as indicated by the 
arrow in the figure (see inset). Therefore, by only considering the adsorption and
desorption  isotherms, one would  predict the occurence of a first-order phase
transition. (By naively integrating the Gibbs adsorption relation
along the adsorption branch from $\mu=-\infty$ to a sufficiently large value,
say $\mu=-4.5$,  and along the
desorption branch from $\mu=-4.5$ to $\mu=-\infty$ (see, e.g.,
Ref.\cite{STC1999}), one  would predict a crossing
at another value of $\mu$; but this
procedure cannot be used, as explained previously.) Secondly, a close inspection of
the results  in the region of hysteresis shows that the states
that yield the absolute minimum of $\Omega$
do not belong to the adsorption or desorption branches (as can be seen
in the inset of Fig. 7).
 This means that the true equilibrium isotherm
is somewhere in between. This is  illustrated in Fig. 8 where we plot
the reduced grand potential $\Omega^*=\Omega/w_{ff}$ versus  $\rho_f^0$, the uniform  seed  density
used to start the iteration procedure at constant chemical
potential (second protocal described in section II B). In this case, the lowest value of the grand potential, $\Omega^{eq}$, is
obtained for $\rho_f^0=0.56$ and the associated value of the average
fluid density is $\rho_f^{eq}=0.399$, whereas one
has $\rho_f=0.352$ and $\rho_f=0.709$ on the adsorption and desorption
branches, respectively. (Note that there are no other solutions than the
extremal ones when $\rho_f^0<0.34$ and $\rho_f^0>0.86$.)

The same  study, performed for each value of $\mu$, yields the 
equilibrium isotherm shown in Fig. 9. The curve is not smooth (there
is a series of small jumps), but it is quite different from 
the pseudo-van der Waals, discontinuous isotherm constructed from the adsorption
and desorption branches. 

Since our search of  local minima
 is not exhaustive, it is important to check that we still get a good approximation of the true
equilibrium isotherm.  For instance, for some matrix realizations the calculated isotherm
is not  a monotonously increasing function of $\mu$ (i.e., $\partial\rho_f^{eq}/ \partial\mu <0$), even when using a very 
small mesh size for $\rho_f^0$ : this  implies that the equilibrium state cannot
be found just with uniform seed configurations,  and these samples were discarded. 
As a general rule, we find that the Gibbs adsorption equation,
$\rho_f=-\left.\partial \omega / \partial \mu\right|_T$ is very well satisfied  along the isotherm calculated from the locus of
the lowest grand potential values, as shown in Fig. 10. This is in contrast with the behavior
observed along the adsorption and  desorption branches (see preceding
section); contrary to what occurs along these latter, $\Omega^{eq}$ is  a {\it continuous} function of $\mu$ even for
finite-size samples\cite{Rem}. This property of thermodynamic
consistency that must be satisfied by the system at equilibrium
can thus be used to control the success (or the failure) of the seed strategy  for 
searching for the equilibrium state.

For some values of the chemical potential, we have also investigated how
$\Delta \Omega$, the gap in grand potential between the two lowest minima,
varies with the system size.  These calculations, unfortunately, are
computationally very demanding, and our  results are 
too limited to conclude that $\Delta \Omega$ is an extensive quantity
(i.e., of order $N$). We believe that this is an important issue that 
deserves a careful investigation in the near future: it is indeed
possible that the system enters a ``glassy phase'' in some region of the parameter space, with several
local minima contributing to the equilibrium properties in the
thermodynamic limit (then $\Delta \Omega=O(1)$), as is the case  for the RFIM near
criticality\cite{LMP1995} (this would correspond to a 
replica-symmetry breaking mechanism in the replica method). A
convenient and intuitive 
way of taking into account this possibility is to consider a  grand partition function
that is a sum over all the solutions of the mean-field equations
weighted by their associated Boltzmann factor\cite{DY1983}, i.e., $\Omega =-(1/\beta) \ln(\sum_{\alpha}e^{-\beta \Omega^{\alpha}})$. For all the
sytems studied here (and $L$ sufficiently large) 
we have found that the  isotherms calculated from this weighted
mean-field approach are not significantly different
from those obtained by only considering the solution with the lowest grand potential.

Finally, the isotherms obtained by averaging over many different
matrix realizations are shown in Figs. 11 and 12. As usual, the averaging
process smears out all the discontinuities present in the isotherms of
the individual samples. Both the averaged equilibrium isotherm, $[\rho_f^{eq}(\mu)]$,
and the curve obtained by averaging the pseudo-van der Waals isotherms obtained
from the adsorption and desorption branches in each sample are
smooth. (Note that this latter is different from the isotherm that
would be obtained from looking at the crossing of the two averaged curves $[\Omega^{ads}(\mu)]$
and $[\Omega^{des}(\mu)]$: this again would lead to a discontinuity.) As can be seen in Fig. 11, the two curves
are quite distinct, which stresses again the failure of the standard
van der Waals picture. 
The equilibrium isotherm depends on the system size but 
this dependence is weak, as shown in Fig. 12. Moreover, the maximum
value of $\partial [\rho_f^{eq}(\mu)]/\partial \mu$ stays  almost constant (or even slightly
decreases)  as $L$ increases. We thus 
conclude that there is no transition at equilibrium in the infinite system.

\subsection{The weak-disorder regime}

The grand-potential minima for $y=1$ and a matrix of size
$L=48$ are presented in Fig. 13 (this is the same sample as in Fig. 6). As could be expected from the shape of
the hysteresis loop,  the presence of artificial liquid-like
metastable states along the desorption isotherm results in a curve
$\Omega^{des}(\mu)$ that is isolated from the other states and that ends
abruptly. We have previously pointed out that this artefact is due to using
periodic boundary conditions. But the main difference with the case $y=1.5$ concerns 
the states that yield the  absolute
minimum of the grand potential. They are indeed very close to the
adsorption branch when $\Omega^{ads}<\Omega^{des}$ and to the desorption branch
when $\Omega^{des}<\Omega^{ads}$. Accordingly,  as shown in Fig. 14,  one 
observes a large jump  in the equilibrium isotherm at a 
chemical potential $\mu_t$ that is  very close to the one at which $\Omega^{ads}=\Omega^{des}$. 
Since a similar  jump in $\rho_f^{eq}$ is present  in all matrix
realizations, one may ask whether there exits  a genuine  first-order transition  in
the thermodynamic limit. To answer this question, one needs to  perform a finite-size
scaling analysis of the average equilibrium isotherm.  In the present
case, this study is somewhat complicated by the fact that 
the large jump  is  sometimes  accompanied by several 
smaller discontinuities, as is the case for two of  the isotherms shown
in Fig. 15.  This  leads to an average isotherm that has a
rather complex shape, as illustrated by the lower part of the $L=16$
curve in Fig. 16 showing  the average equilibrium isotherms for various
system sizes (we recall that this study is limited to rather small systems  because of
the considerable computational effort required by the search of the
equilibrium states when $N$ is large).  Although the appearance  of
several discontinuities may be a finite-volume artefact, the possible
existence of two or more capillary transitions at different
chemical potentials in the thermodynamic limit has been 
considered in a previous work using the replica
formalism\cite{KRTP1998}. To investigate this  question a 
separate analysis of the different parts of the isotherms is required, and  we defer this
delicate study to a future work.

We thus analyze the curves in Fig. 16 focusing on the transition
associated with the largest  discontinuity in the individual
isotherms. Such a discontinuity in a finite system is of course an
artefact of the mean-field approximation: in an exact theory, one would only observe a maximum
in the susceptibility, $\chi(\mu)=\partial \rho_f^{eq}(\mu)/\partial\mu$ (this is also true with the weighted mean-field theory described
above); then, if this maximum occurs  at
the same $\mu_t(L)$ in all matrix realizations  $x=\{\eta_i\}$ of linear size $L$, the average susceptibility
$\chi(\mu_t)=[\chi(\mu_t,x)]$ would scale as  $L^3$ as is usual with first-order phase
transitions, e.g., $\chi(\mu)\approx L^3 F(L^3(\mu-\mu_t(L))$  around $\mu_t(L)$
where $F$ is a (non-universal) scaling funtion.  The finite-size
scaling behavior of the transition associated with  capillary
condensation in a disordered solid  is however different 
because every specific sample $x$  is characterized by a
different location  $\mu_t(x, L)$ of the maximum of $\chi(\mu,x)$(or, in the
mean-field approximation, a different
location of the discontinuity, as can be seen  in Fig. 15). According to a 
standard  argument first put forward by Brout\cite{B1959}, we expect that all extensive thermodynamic
quantities in a disordered system  far from criticality  are
self-averaging, with a Gaussian probability
distribution around their mean value and  a variance proportional  to
$L^{-3}$. Although $\mu_t(x, L)$ is not the density of an
extensive quantity,  its value is indirectly determined by the
fluctuations of the local random fields, and it seems reasonable (for $L$ large enough and
far enough from a critical point) to
assume that it also fluctuates around a mean value $\mu_t(L)=[\mu_t(x,L)]$ with a
variance  $\delta\mu_t(L)^2\propto L^{-3}$. This sample-to-sample fluctuation of
$\mu_t(x, L)$ implies that at a first-order transition the maximum of the average
susceptibility should scale as $L^{3/2}$ instead of
$L^3$. Specifically, we expect that $\chi(\mu)=\partial [\rho_f^{eq}(\mu,x)]/\partial\mu\approx
L^{3/2} f(L^{3/2}\{\mu-\mu_t(L)\})$ around $\mu_t(L)$, where $f$ is some (non-universal)
scaling function. Our DFT calculations  for very weak
disorder (e. g., $\rho_m=0.1$ and $y=0.6$) strongly support this scaling ansatz, and, as shown in Fig. 17, one
can also reach a reasonably good
collapse of the equilibrium isotherms for $y=1$, $\rho_m=0.25$, and
$T^*=0.8$  using the
scaling reduced variable $L^{3/2}\{\mu-\mu_t(L)\}/\mu_t(L)$.  The curves in
Fig. 16 can also be fitted  by the function $[\rho_f^{eq}(\mu)]=a
\tanh(bL^{3/2}\{\mu^*-\mu_t^*(L)\})+c$ where $a,b$ and $c$ are adjustable
constants (in both methods, however, there are some deviations for 
$\mu^*<\mu_t^*(L)$ which we attribute to the presence of  the additional discontinuities
in the individual  isotherms). 
We thus conclude from this study that  a genuine  first-order equilibrium
transition occurs in the thermodynamic limit for $y=1$, $\rho_m=0.25$, and
$T^*=0.8$.

Rather remarkably, this transition is not accompanied by a discontinuous
behavior on the adsorption branch. Indeed, as can be seen from
Fig. 18,  for $L$ sufficiently large, the curves collapse onto a clearly continuous isotherm. (Note that we had to
consider larger systems than for the equilibrium case in order to reach the asymptotic regime.) The
fact that a continuous (but out-of-equilibrium) filling process does not imply the absence of a
sharply defined  first-order (equilibrium) transition cannot be explained  in the
traditional picture of capillary condensation based on the independent pore model and is at odds
with the common lore in the adsorption community.

\section{Conclusion}

The main conclusion of this study is that most of the 
phenomenology  of capillary condensation  in
disordered porous solids  can be reproduced by a theoretical model
that focuses on the properties of a free-energy (or grand
potential) landscape with  many local metastable states.  Thermal
fluctuations,  which  are neglected in the treatment, do not seem to play an important
role in  usual adsorption experiments, and it  is the evolution of the
landscape with the gas pressure in the reservoir (or the chemical potential), the  
temperature, and  the amount of disorder (indirectly controlled by the
solid porosity and its wettability) that explains the changes  in the
hysteresis loops and the scanning curves.  Fluids in disordered solids
are indeed (complicated) experimental realizations of random-field systems for which it
is well known that disorder may lead to diverging barriers to
relaxation as $T\to 0$\cite{FGK1988}. Such externally-driven systems have been extensively studied in recent
years at $T=0$  (see, e.g., Refs.\cite{S1993,VP1994}), and the present
work based on the mean-field density
functional theory can be viewed as an extension of such approaches to finite
temperatures.  In these systems, hysteresis  is not the necessary signature of
an underlying  equilibrium  phase transition: this feature is illustrated  by the
calculations of section  IV.B, but is still not well accepted in the
adsorption community.  As is discussed
elsewhere\cite{KRT2002},  hysteresis in capillary condensation can
also be accompanied by  {\it out-of-equilibrium} phase transitions whose nature depend crucially on the
presence of the external interface between the solid and the gas reservoir.

%\end{thebibliography}

\newpage

\begin{figure}
\begin{center}
\resizebox{10.5cm}{!}{\includegraphics{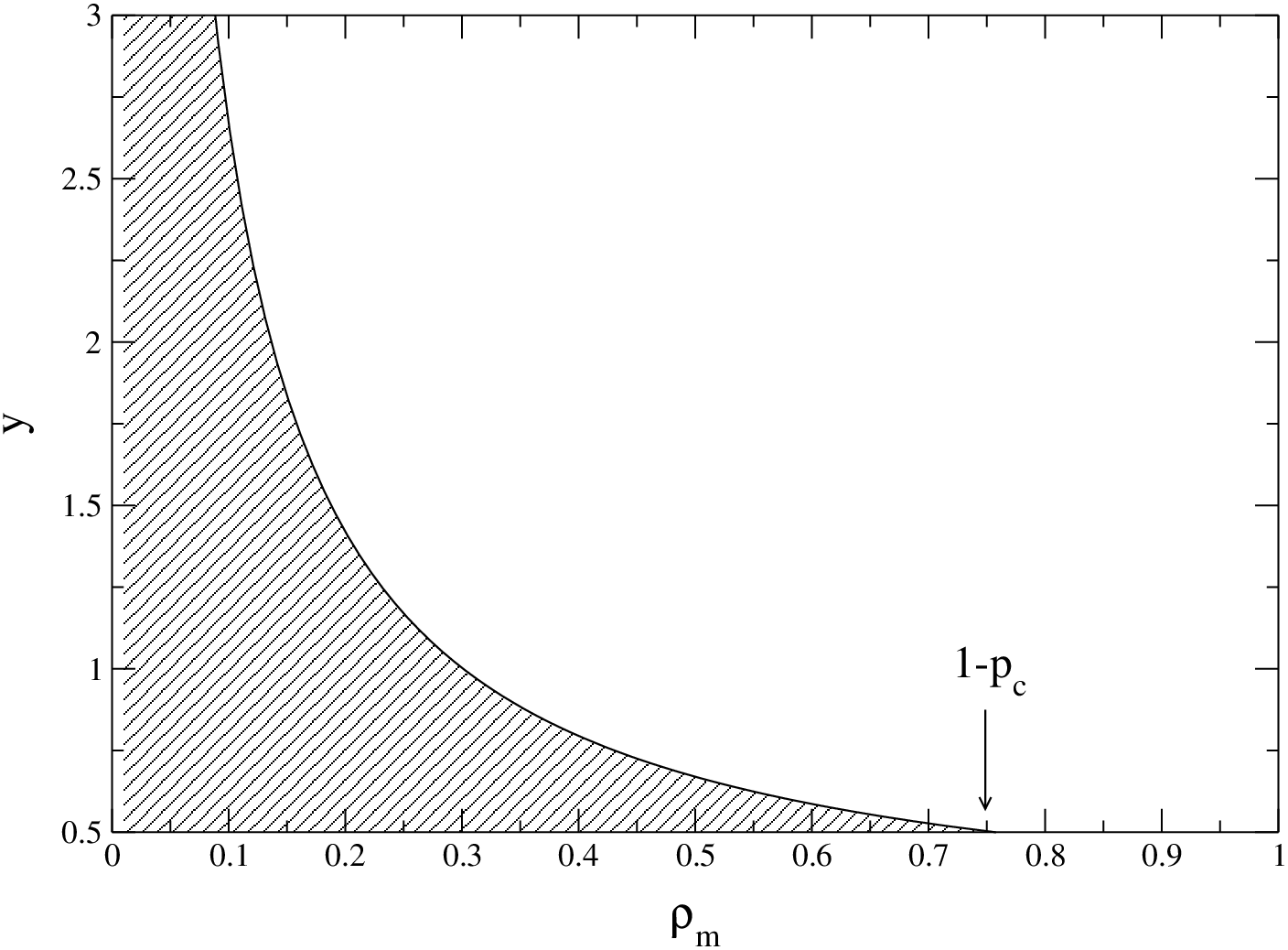}}
%\vspace{.4cm}

\caption{Putative  equilibrium phase diagram of the model in the $\rho_m-y$
plane for $y\geq 1/2$. A genuine capillary condensation occurs  for
$T<T_{cc}(\rho_m,y)$ in the hatched region  with $T_{cc}(\rho_m,y)=0$ 
at the boundary. For $\rho_m>1-p_c$, there is no phase
transition because the void space ceases to percolate.}

\end{center}
\end{figure}

\begin{figure}
\begin{center}

\resizebox{10.5cm}{!}{\includegraphics{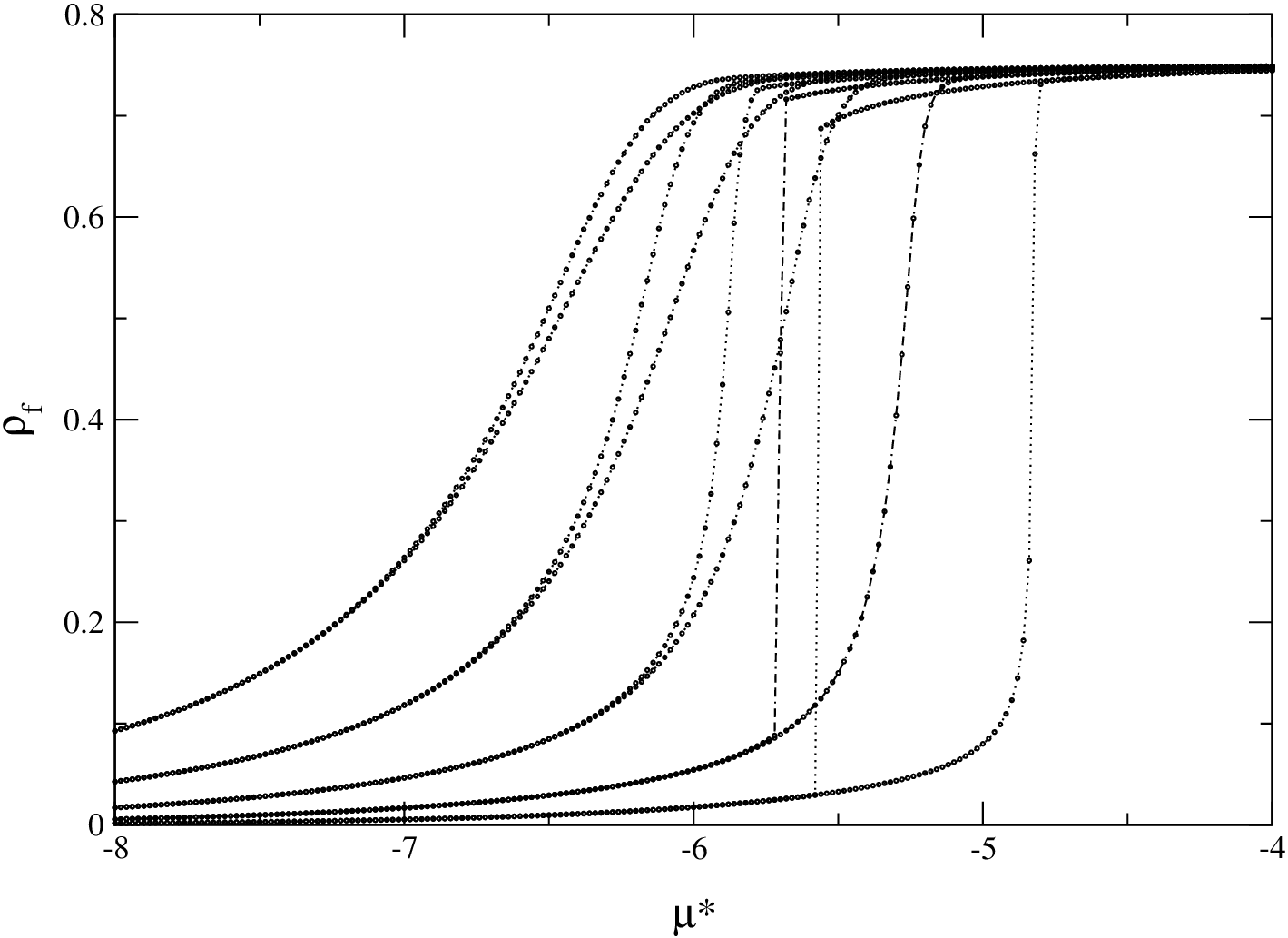}
}%\vspace{.4cm}

\caption{DFT prediction for the sorption isotherms, $\rho_f$ versus $\mu*=\mu/w_{ff}$, for $T^*=0.8$
and  $\rho_m=0.25$. From  left  to right: $y=2$,  $1.75$,
$1.50$, $1.25$, and $1$. The  results have been averaged  over
$10$ matrix samples of linear  size $L=48$. Recall that at complete
filling, $\rho_f=1-\rho_m=0.75$. }
\end{center}
\end{figure}

\newpage
\begin{figure}
\begin{center}
\resizebox{10.5cm}{!}{\includegraphics{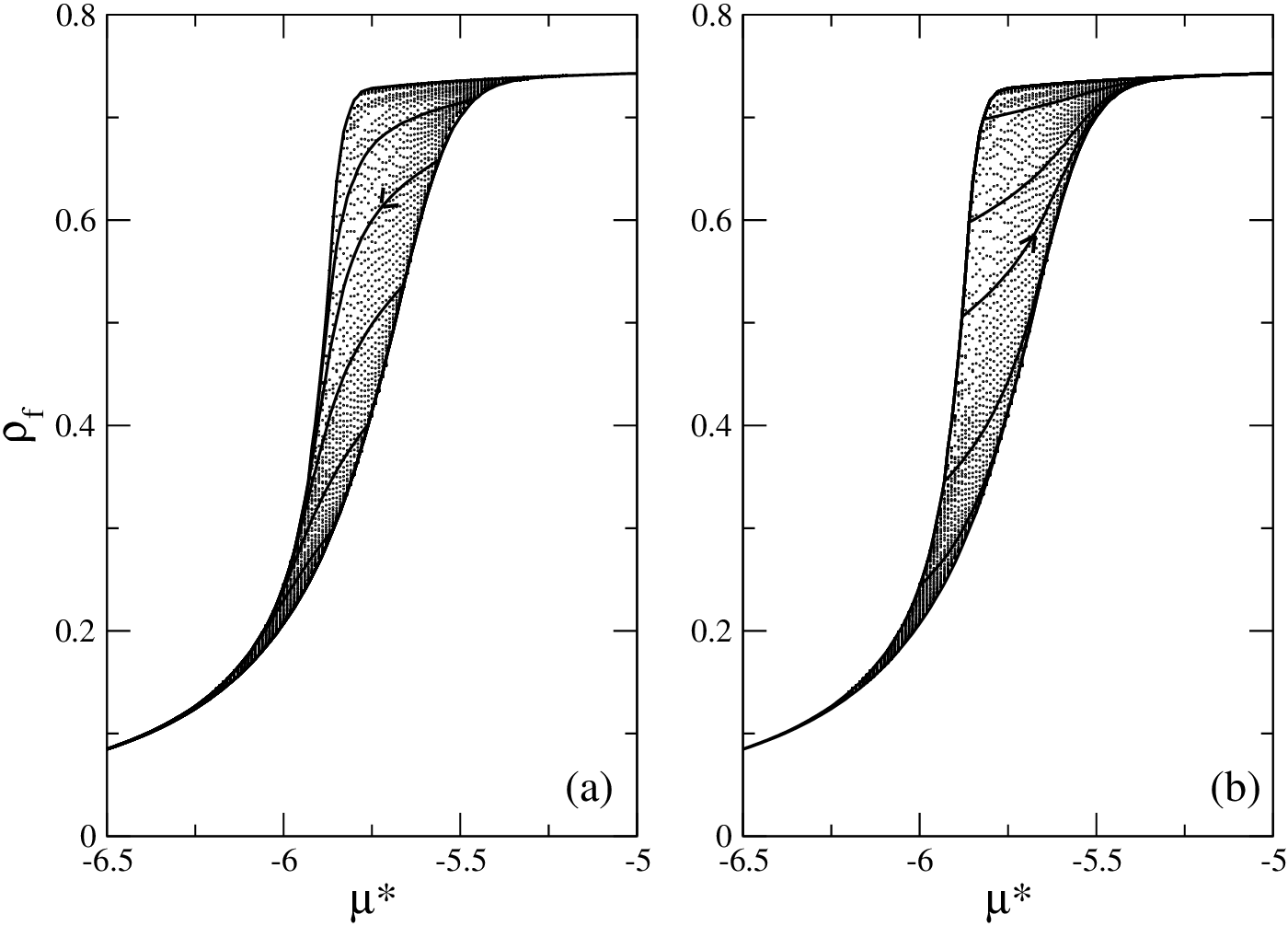}}
%\vspace{.4cm}
\caption{DFT predictions for the desorption (a) and  adsorption (b)
scanning curves for $y=1.5$ in a matrix of
linear size $L=48$. Also shown are the many metastable states obtained
by solving Eqs. (4) according to the second protocol described in the text.}
\end{center}
\end{figure}

\begin{figure}
\begin{center}
\resizebox{10.5cm}{!}{\includegraphics{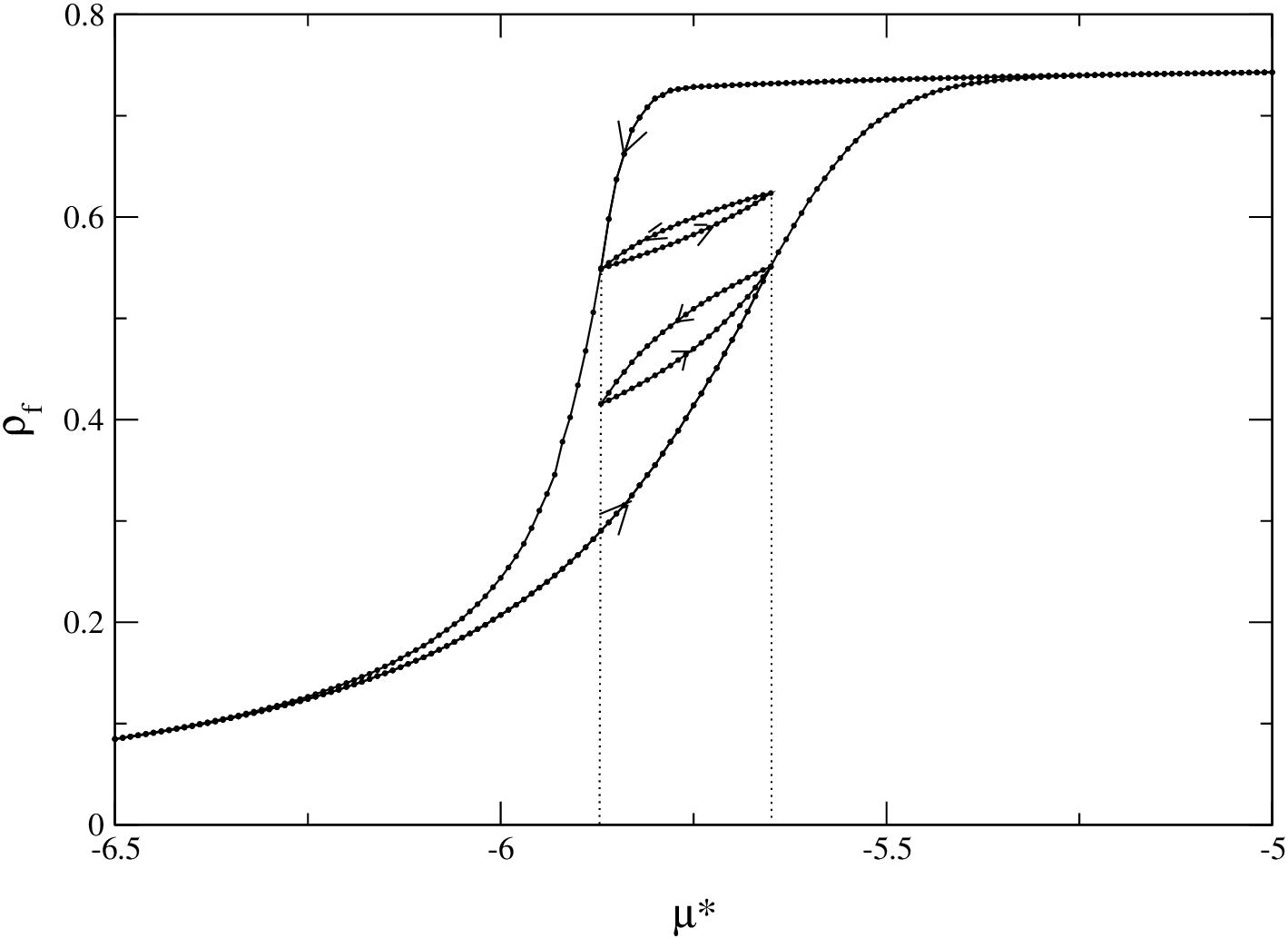}}%\vspace{.4cm}

\caption{DFT predictions for the global  hysteresis loop and two
representative subloops for $y=1.5$. The arrows denote the
different  filling-draining cycles. The results have been averaged
over 10 matrix samples of linear size $L=48$.}
\end{center}
\end{figure}

\begin{figure}
\begin{center}
\resizebox{10.5cm}{!}{\includegraphics{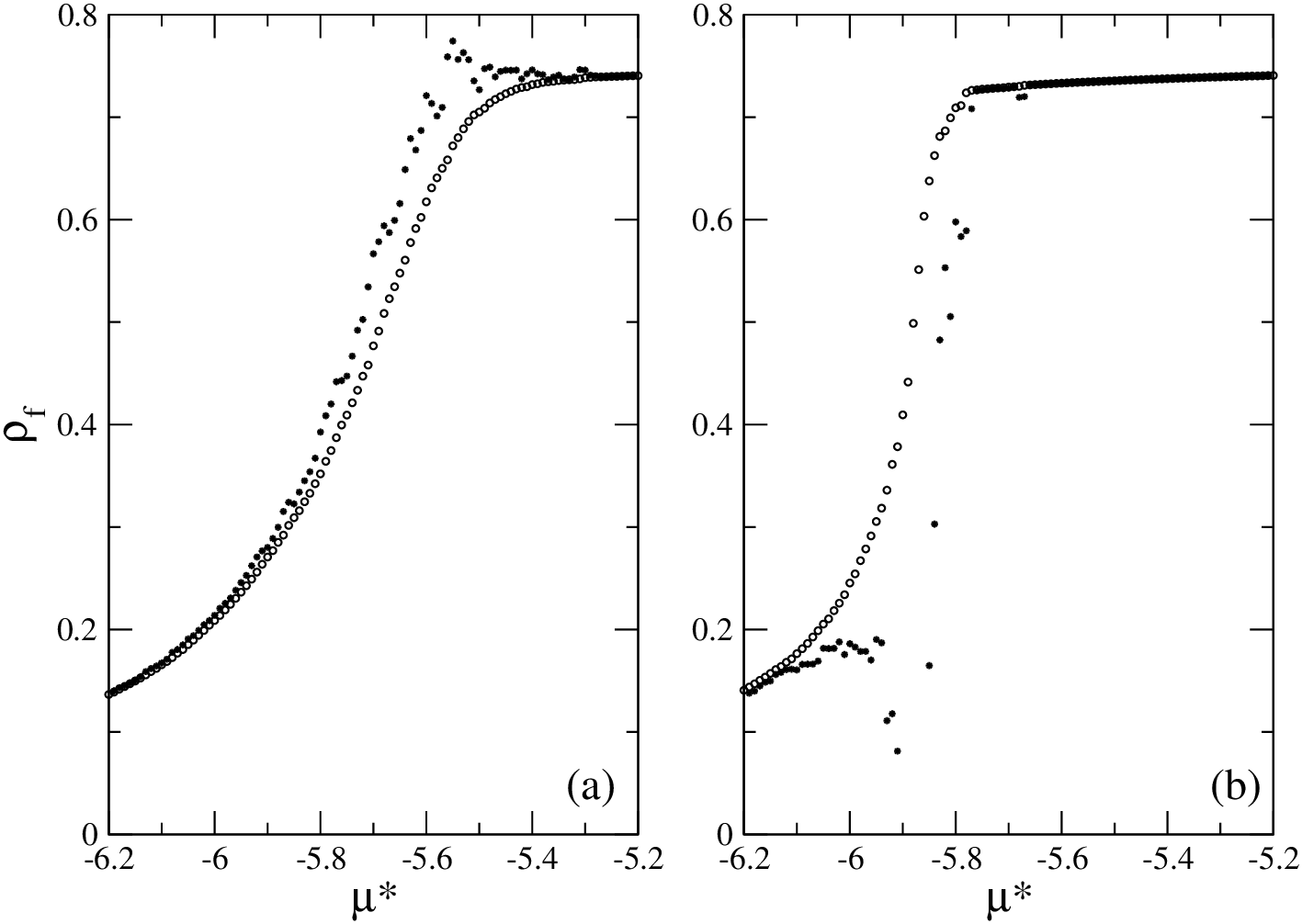}}%\vspace{.4cm}
\caption{Check of thermodynamic consistency along the adsorption (a)
and desorption (b)
isotherms for $y=1.5$ ($L=48$). Circles: average fluid
density obtained from the solution of Eqs. (4). Stars: related quantity
obtained by differentiating the corresponding grand potential.}
\end{center}
\end{figure}

\begin{figure}
\begin{center}
\resizebox{10.5cm}{!}{\includegraphics{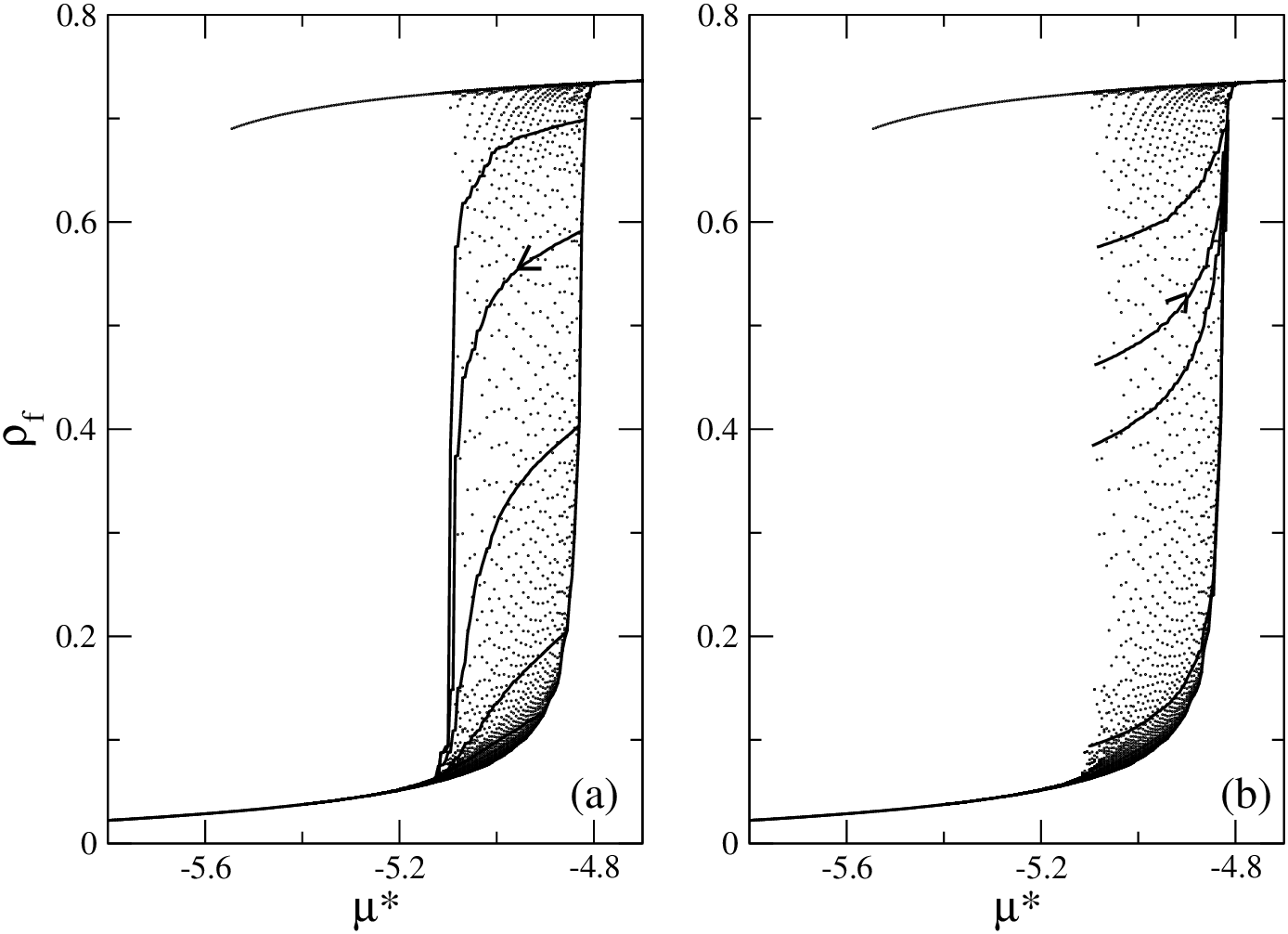}}%\vspace{.4cm
\caption{Same as Fig. 3 for $y=1$}
\end{center}
\end{figure}

\begin{figure}
\begin{center}

\resizebox{10.5cm}{!}{\includegraphics{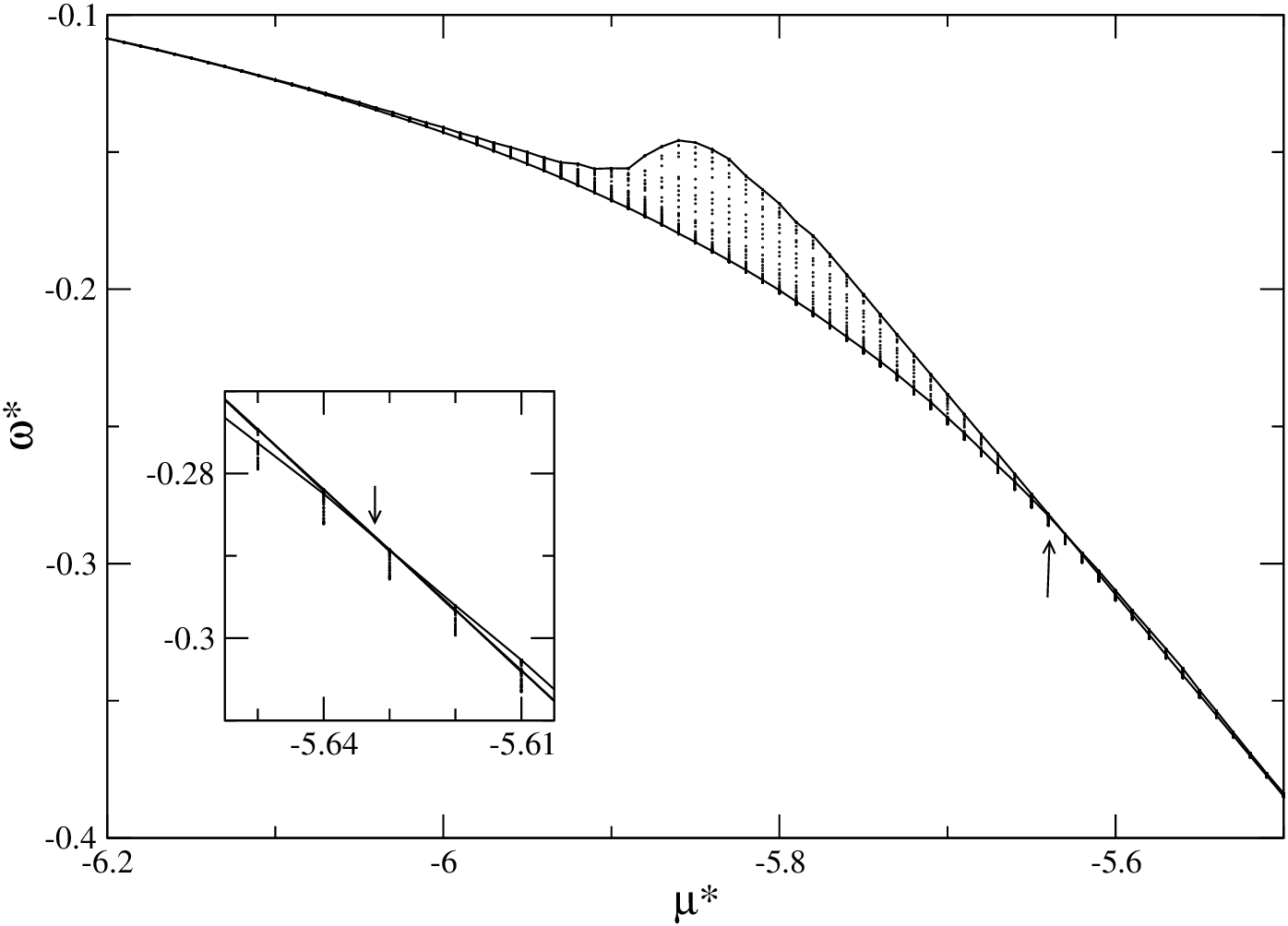}
}%\vspace{.4cm}

\caption{Reduced grand-potential  density, $\omega^*=\Omega/(Nw_{ff})$, associated with the solutions of the
mean-field equations for $y=1.5$ ($L=48$). The arrow
indicates the crossing of the adsorption and desorption branches and
the inset shows a zoom up of this region.}
\end{center}
\end{figure}

\begin{figure}
\begin{center}
\resizebox{10.5cm}{!}{\includegraphics{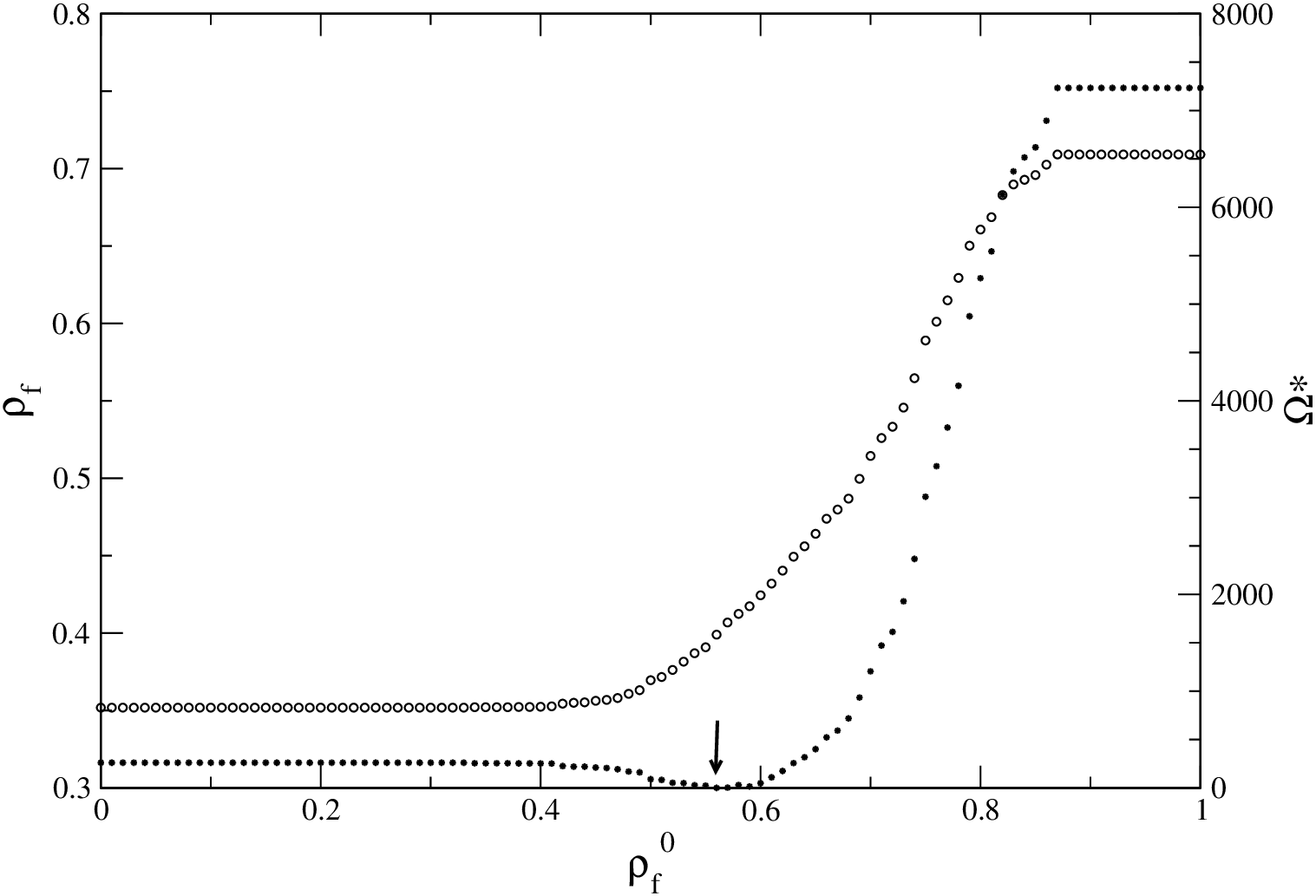}}%\vspace{.4cm}
\caption{Average fluid density $\rho_f$ (circles) and
reduced grand-potential $\Omega^*$ (stars) as functions of the initial seed
density $\rho_f^0$ for $y=1.5$ and $\mu^*=-5.80$ ($L=48$). Here $\Omega$ is measured with
respect to its minimum value, $\Omega^{eq}$, obtained for $\rho_f^0=0.56$ (arrow).}
\end{center}
\end{figure}

\begin{figure}
\begin{center}
\resizebox{10.5cm}{!}{\includegraphics{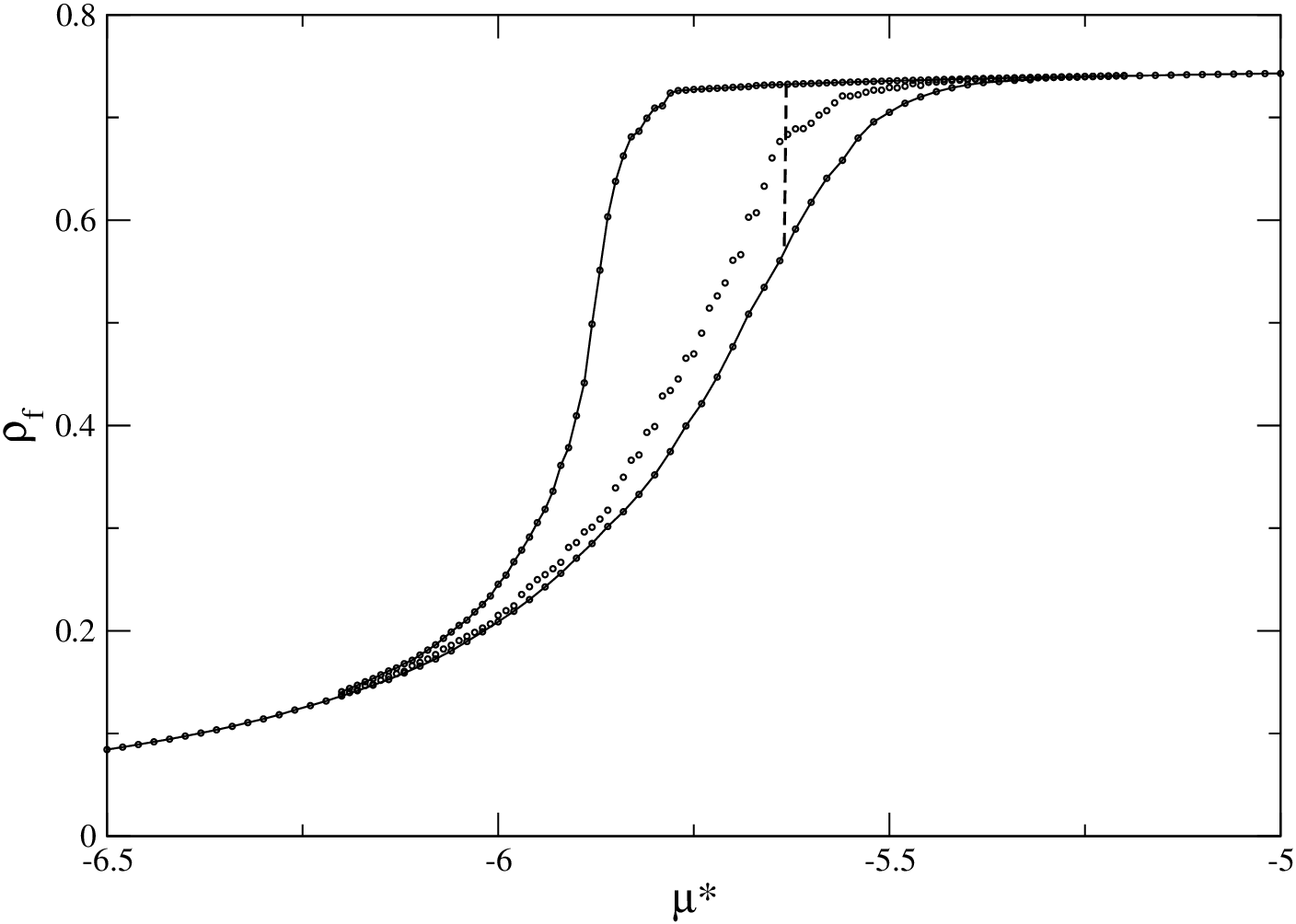}}%\vspace{.4cm}

\caption{Hysteresis loop and 
equilibrium isotherm for  $y=1.5$ ($L=48$). The dashed
line indicates the first-order transition predicted by considering only  the
adsorption and desorption isotherms.}
\end{center}
\end{figure}

\begin{figure}
\begin{center}
\resizebox{10.5cm}{!}{\includegraphics{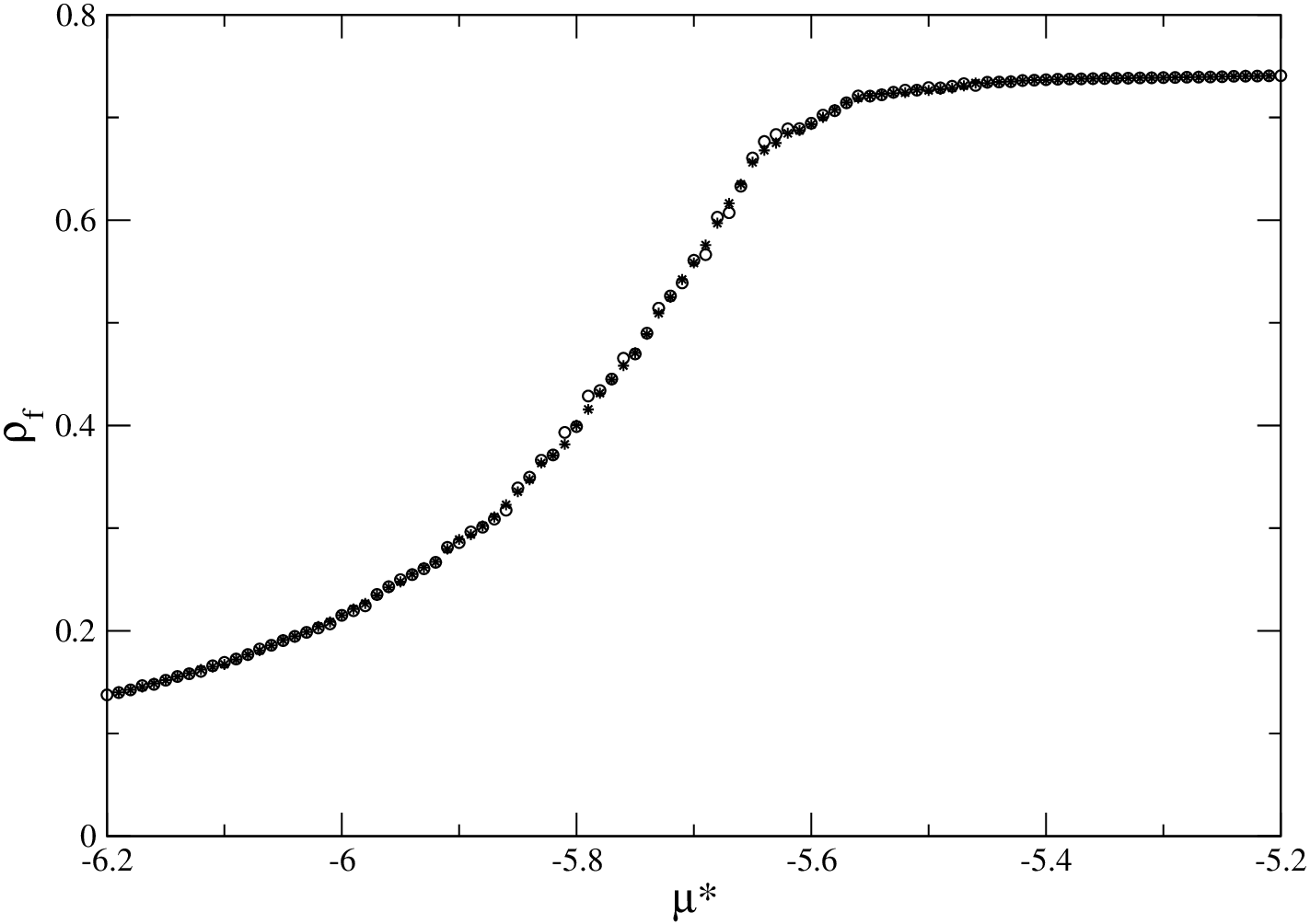}}%\vspace{.4cm}

\caption{Check of thermodynamic consistency along the equilibrium
isotherm for $y=1.5$ ($L=48$). Circles: average fluid
density obtained from the solution of Eqs. (4). Stars: quantity
obtained by differentiating the corresponding grand-potential.}
\end{center}
\end{figure}

\begin{figure}
\begin{center}

\resizebox{10.5cm}{!}{\includegraphics{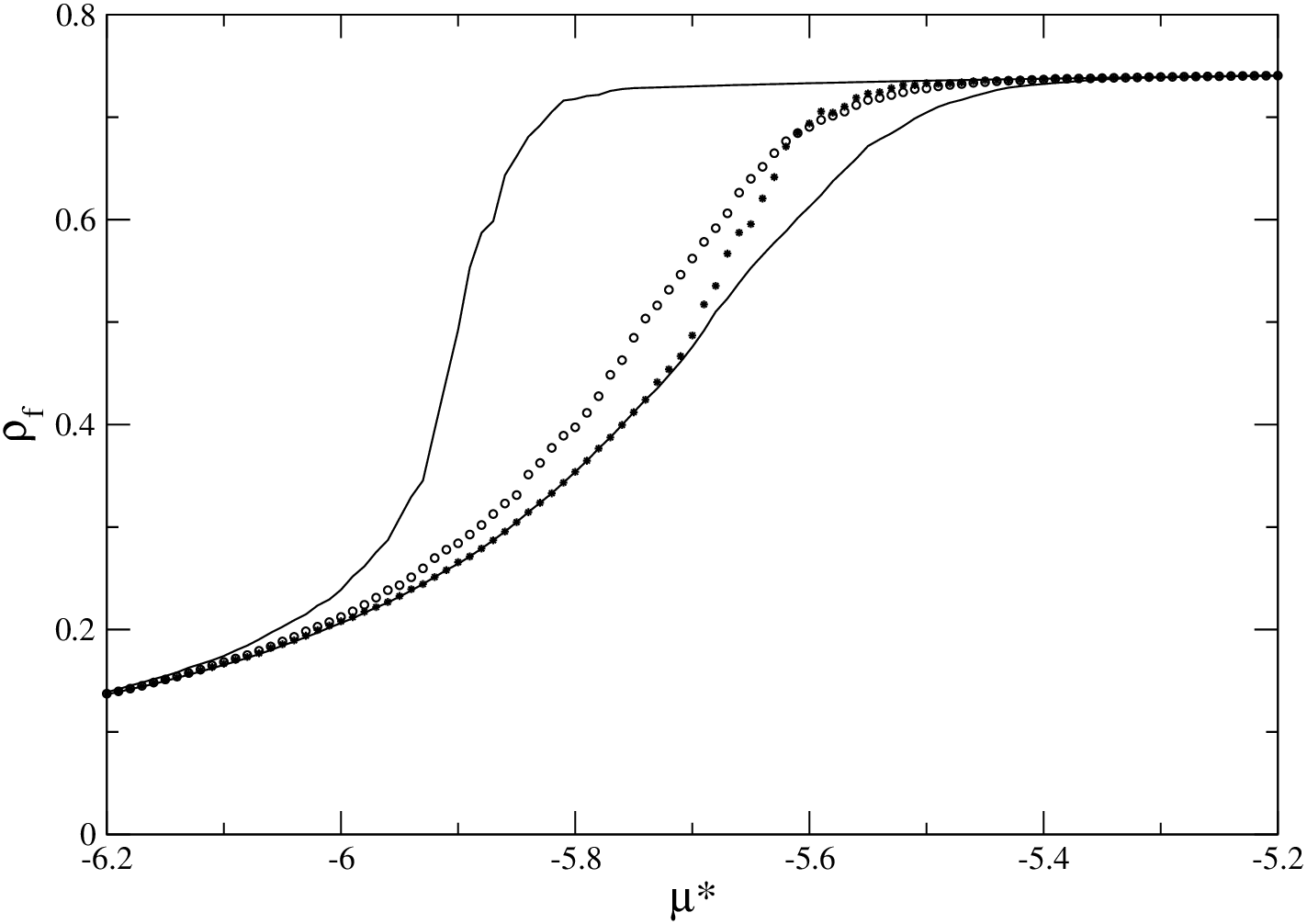}
}%\vspace{.4cm}
\caption{Hysteresis loop,
equilibrium isotherm (circles) and  pseudo-van der Waals isotherm
(stars) for $y=1.5$ in a matrix of linear size $L=16$ . An average over $50$
matrix realizations has been performed.}
\end{center}
\end{figure}

\newpage
\begin{figure}
\begin{center}

\resizebox{10.5cm}{!}{\includegraphics{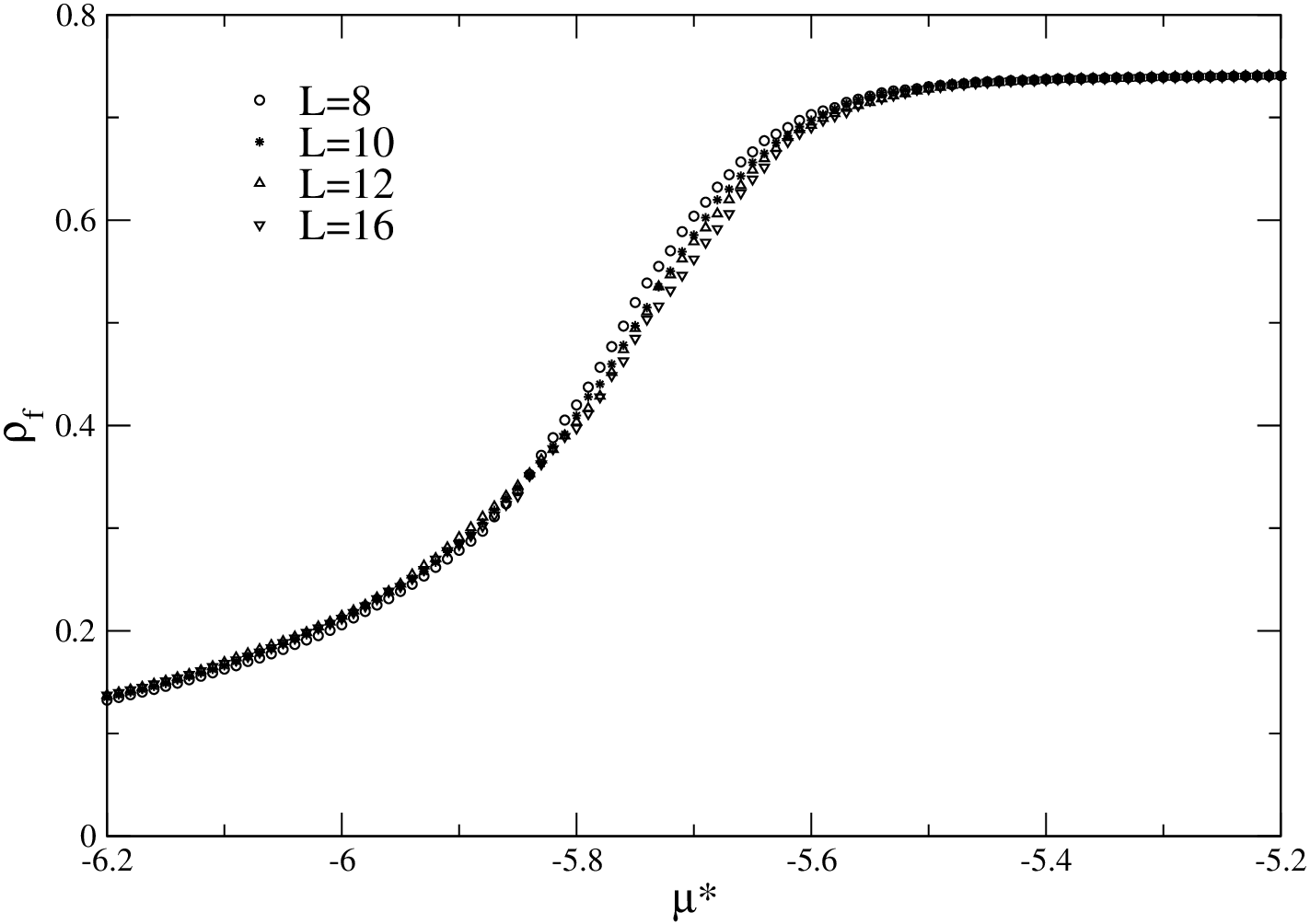}}%\vspace{.4cm}
\caption{Equilibrium isotherms for $y=1.5$ and different
system sizes. An average over $50$ $(L=16)$ to $400$ $(L=8)$ matrix
realizations has been performed.}
\end{center}
\end{figure}

\begin{figure}
\begin{center}

\resizebox{10.5cm}{!}{\includegraphics{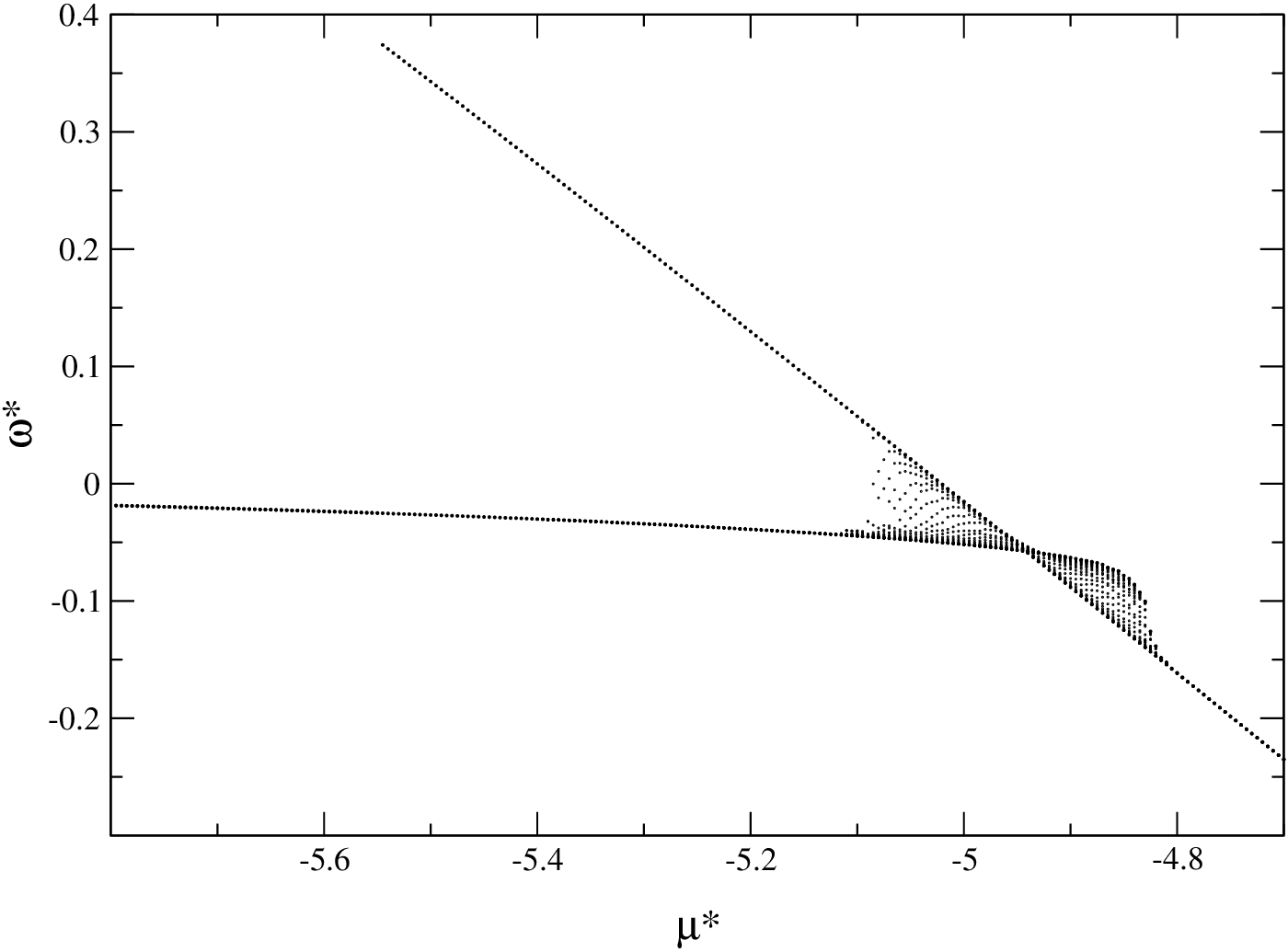}
}%\vspace{.4cm}
\caption{ Reduced grand-potential density associated with the solutions of the
mean-field equations for $y=1$ ($L=48$).}
\end{center}
\end{figure}

\begin{figure}
\begin{center}

\resizebox{10.5cm}{!}{\includegraphics{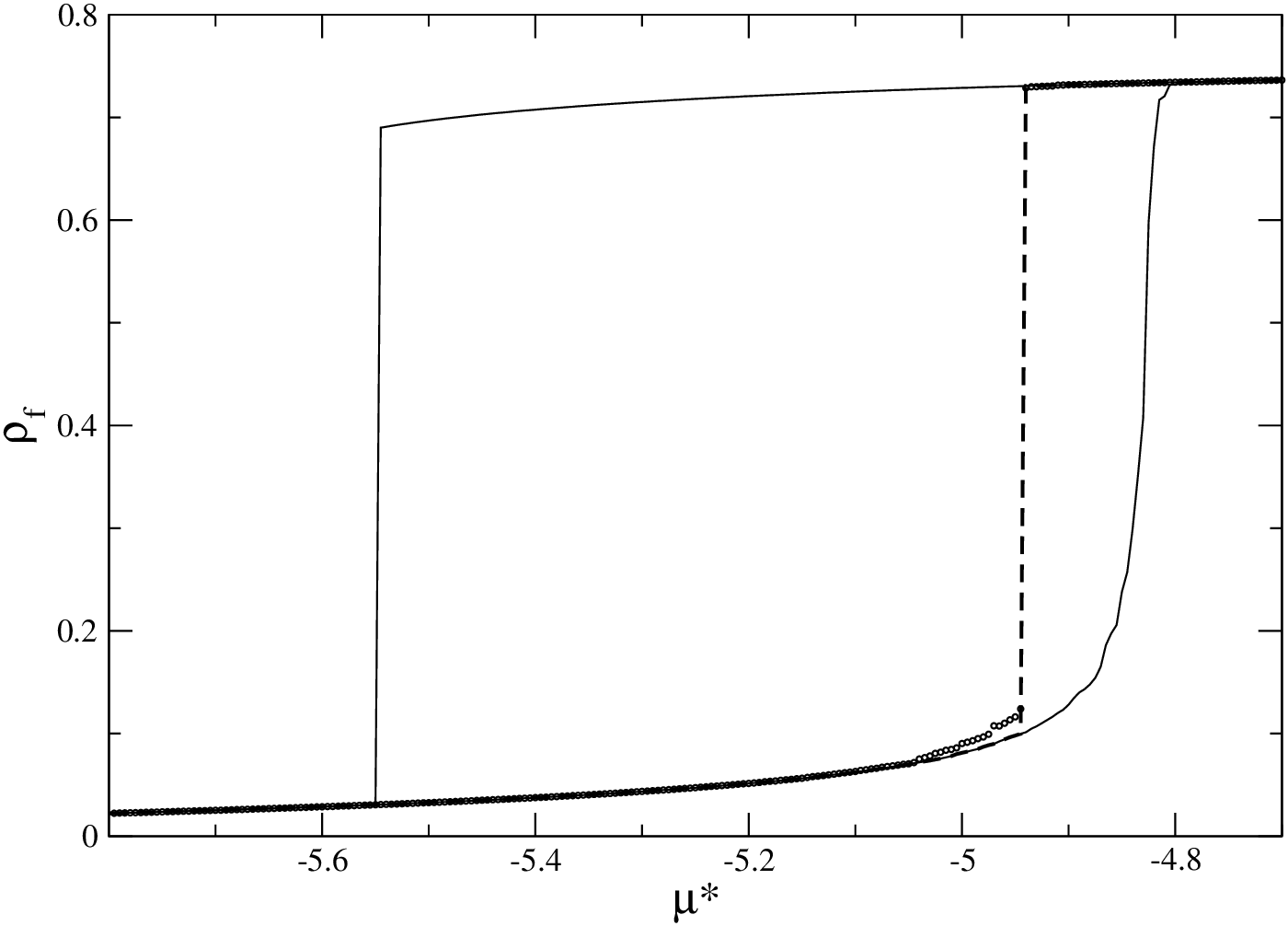}}%\vspace{.4cm}

\caption{Hysteresis loop and 
equilibrium isotherm for  $y=1$  ($L=48$). The dashed
line indicates the first-order transition predicted by considering only  the
adsorption and desorption isotherms.}
\end{center}
\end{figure}

\begin{figure}
\begin{center}
\resizebox{10.5cm}{!}{\includegraphics{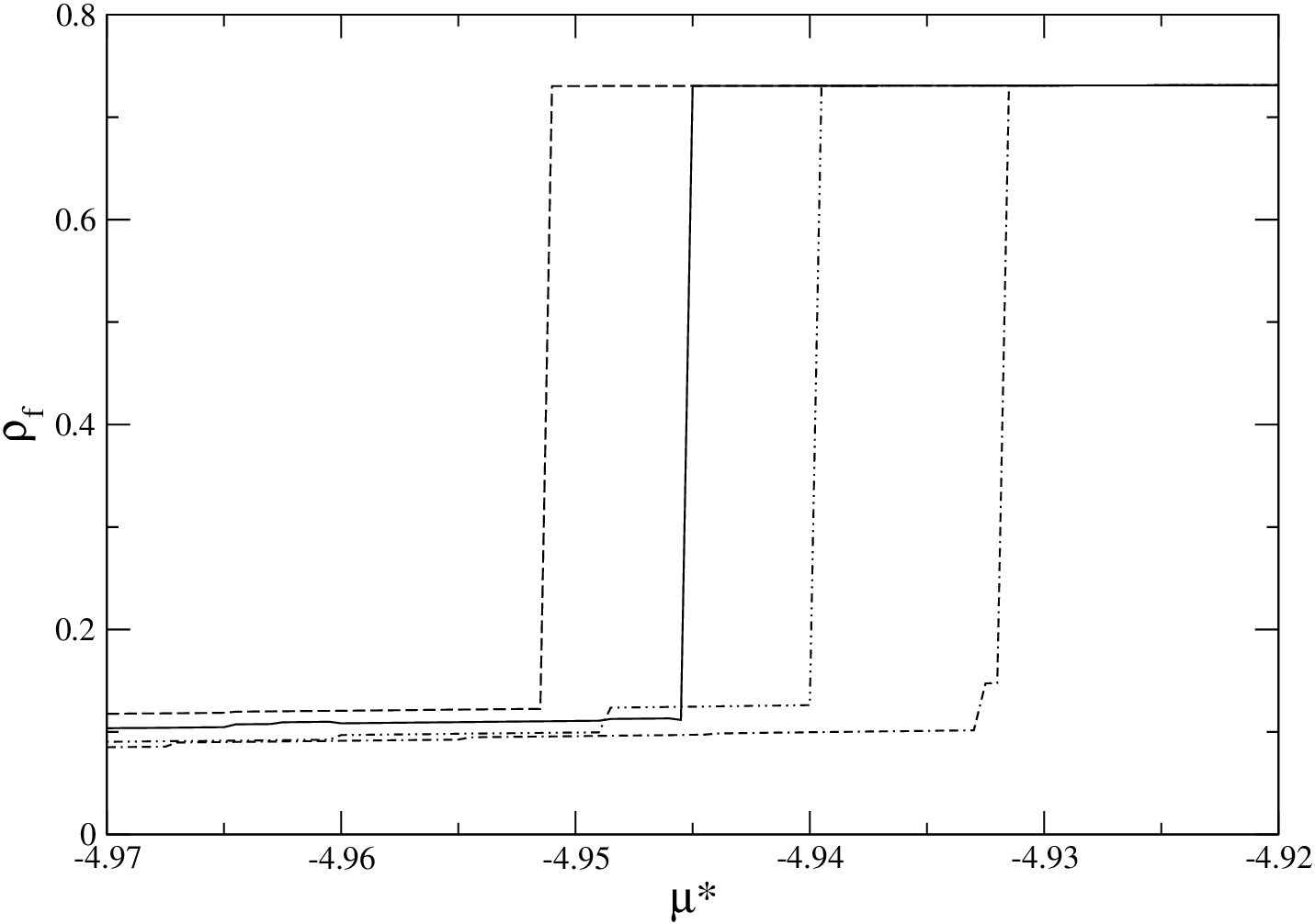}}%\vspace{.4cm}
\caption{Equilibrium isotherms for $y=1$ and 4 matrix
realizations of linear size $L=16$.}
\end{center}
\end{figure}

\begin{figure}
\begin{center}
\resizebox{10.5cm}{!}{\includegraphics{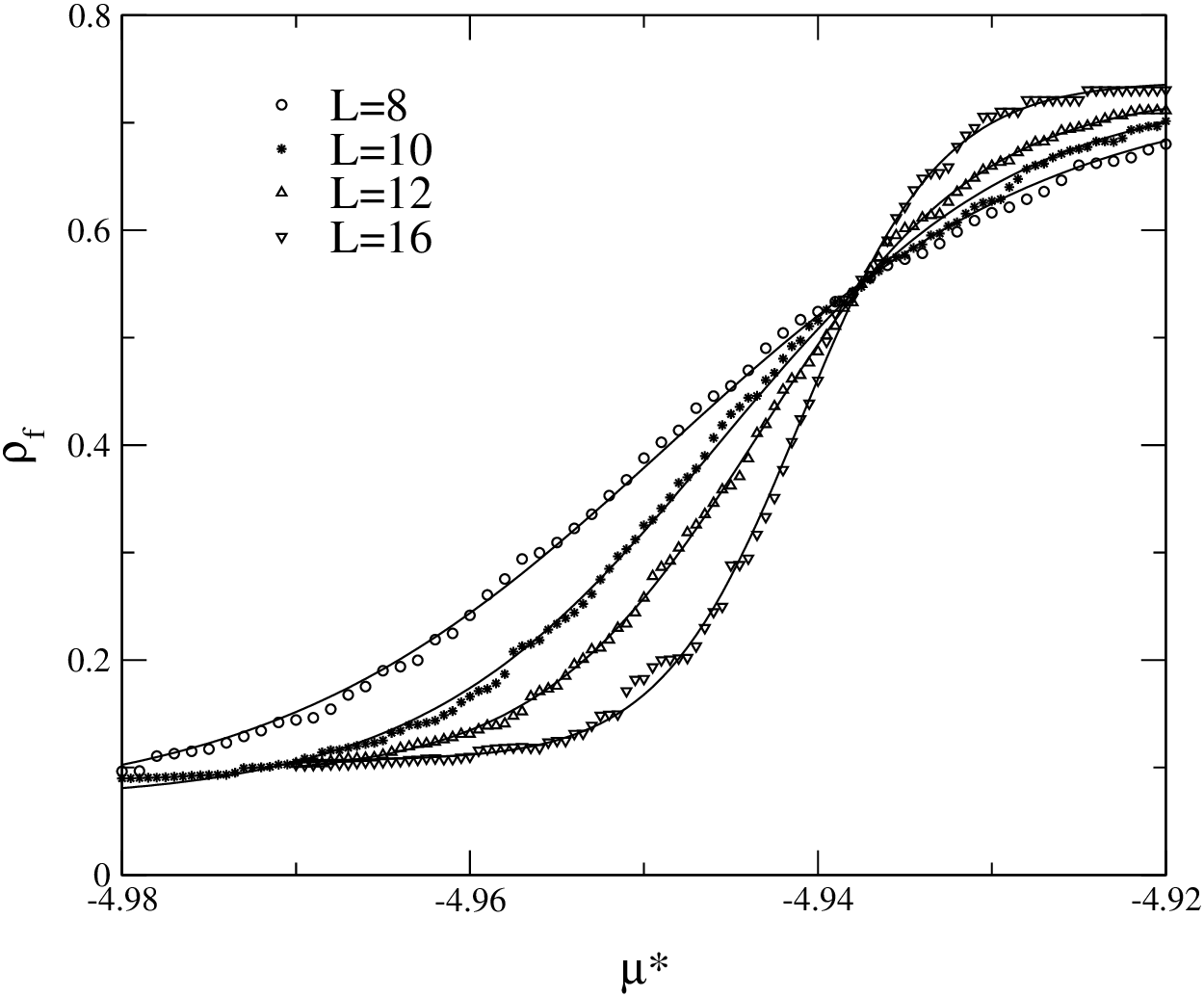}}%\vspace{.4cm}
\caption{Equilibrium isotherms for $y=1$ and different
system sizes. An average over $125$ $(L=16)$ to $400$ $(L=8)$ matrix
realizations has been performed. The solid lines indicate the fit
discussed in the text.}
\end{center}
\end{figure}

\begin{figure}
\begin{center}
\resizebox{10.5cm}{!}{\includegraphics{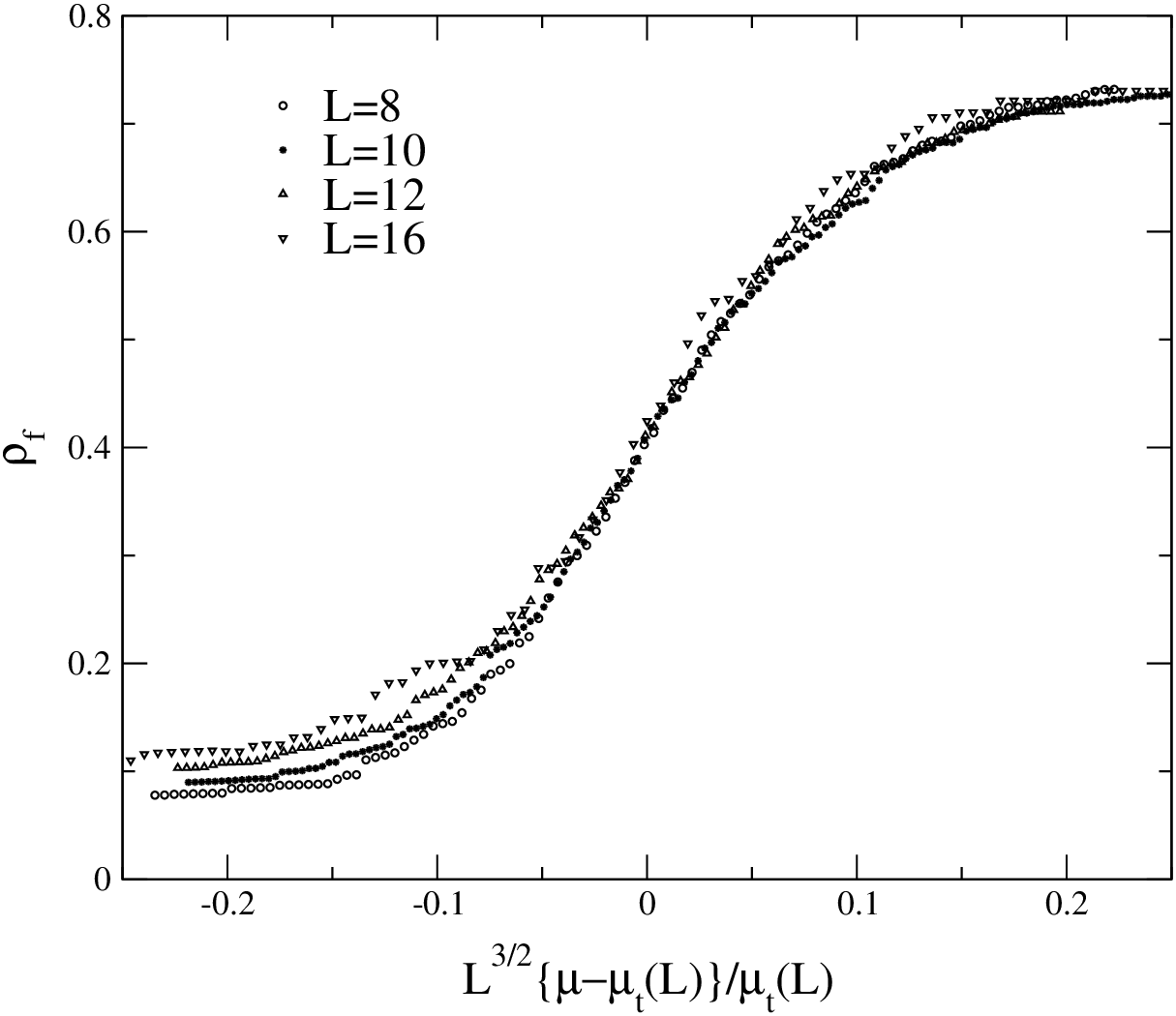}}%\vspace{.4cm}

\caption{Scaling plot of the average equilibrium isotherms shown in
Fig. 16.  $\mu_t(L)$ is the chemical potential that corresponds to the
maximum of the average susceptibility for the system of size $L$.}
\end{center}
\end{figure}

\begin{figure}
\begin{center}
\resizebox{10.5cm}{!}{\includegraphics{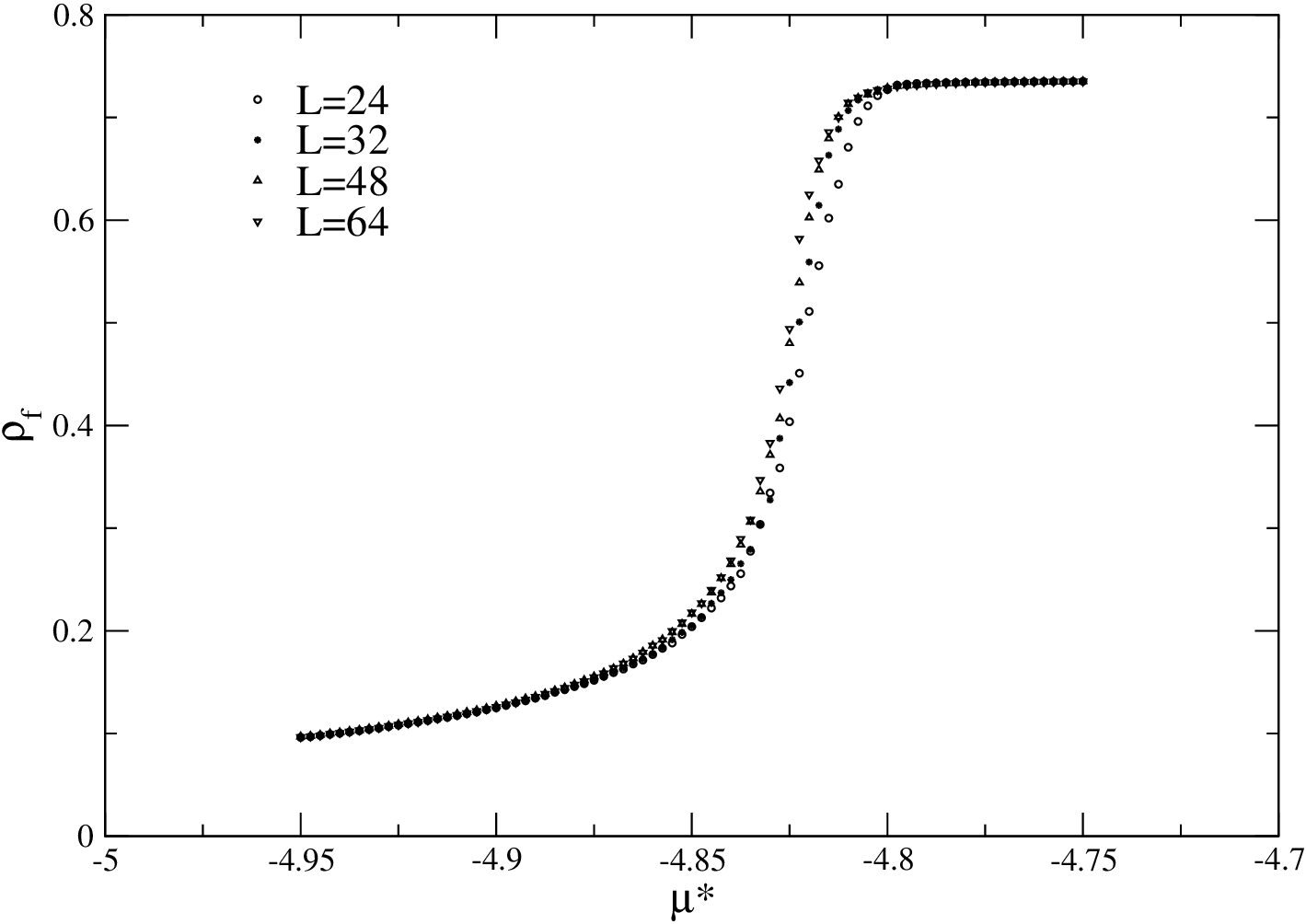}}

\caption{Adsorption isotherm for $y=1.0$ and different
system sizes. An average over $50$  $(L=64)$ to $250$ $(L=24)$ matrix
realizations has been performed.}  
\end{center}
\end{figure}


\begin{references}
%\begin{thebibliography}

\bibitem[*]{AAAuth} The Laboratoire de Physique Th{\'e}orique des Liquides is the UMR 7600 of the CNRS.
\bibitem{E1990} R. Evans, J. Phys.: Condens. Matter \textbf{2}, 8989 (1990).
\bibitem{GGRS1999} For a recent review, see L.D.  Gelb, K. E. Gubbins,
R. Radhakrishnan, and M. Sliwinska-Bartkowiak, 
Rep. Prog. Phys. \textbf{62}, 1573 (1999).
\bibitem{WC1990}A. Wong and M. Chan, Phys. Rev. Lett. \textbf{65}, 2567
(1990); A. Wong, S. B. Kim, W. I. Goldburg, and M. H. W. Chan,{\it et al.},
Phys. Rev. Lett. \textbf{70}, 954 (1993); D. J. Tulimieri, J. Yoon,
and M. H. W. Chan, Phys. Rev. Lett. {\bf 82}, 121 (1999); 
L. B. Lurio {\it et al}, J. Low Temp. Phys. {\bf 121}, 591 (2000).
\bibitem{Z1996} Z. Zhuang, A. G. Casielles, and D. S. Cannell,
Phys. Rev. Lett. {\bf 77}, 2969 (1996).
\bibitem{B1963} A. J. Brown, Ph.D. Thesis, University of Bristol,
1963; some of Brown's results are reproduced by  G. Mason in
Proc. R. Soc. Lond.  A {\bf 415}, 453 (1988), and by
P.C. Ball and R. Evans \cite{BE1989}.
\bibitem{GS1982} S. J. Gregg and K. S. W. Sing, {\it Adsorption,
Surface Area and Porosity} (Academic Press, London,1982).
\bibitem{BE1989} P.C. Ball and R. Evans, Langmuir \textbf{5}, 714
(1989).
\bibitem{E1967}D. H. Everett, in \emph{The Solid Gas Interface},
E. A. Flood Ed. (Marcel Dekker, New York, 1967), vol.  2, p.  1055.
\bibitem{LH2001} M. P. Lilly and R. B. Hallock, Phys. Rev. {\bf B 63}, 174503 (2001).
\bibitem{P1935} F. Preisach, Z. Physik  {\bf 94}, 277 (1935).
\bibitem{PRST1995} E. Pitard, M. L. Rosinberg, G. Stell, and
G. Tarjus, Phys. Rev. Lett. {\bf 74}, 4361 (1995).
\bibitem{M1982} G. C. Wall and R. J. C. Brown, J. Colloid Interface
Sci. \textbf{82}, 141 (1981); G. Mason, J. Colloid Interface
Sci. \textbf{88}, 36 (1982); Proc. R. Soc. Lond. A \textbf{390}, 47 (1983);
A. V. Neimark, Sov. Phys. Tech. Phys. \textbf{31}, 1338 (1986);  M. Parlar and
Y. C. Yortsos, J. Colloid Interface Sci. \textbf{124}, 162 (1988);
N. A. Seaton, Chem. Eng. Sci.,\textbf{46}, 1895 (1991); R. A. Guyer
and K. R. McKall, Phys. Rev. \textbf{B 54}, 18 (1996).
\bibitem{KRTV2001} E. Kierlik, M.L. Rosinberg, G. Tarjus, and P. Viot,
Phys. Chem. Chem. Phys. \textbf{3}, 1201 (2001).
\bibitem{KMRST2001} E. Kierlik, P. A. Monson, M. L. Rosinberg,
L. Sarkisov, and G. Tarjus, Phys. Rev. Lett. \textbf{87}, 055701
(2001).
\bibitem{SM2001} L. Sarkisov, and P. A. Monson, Phys. Rev. E \textbf{65}, 011202 (2001).
\bibitem{WM2002} H.-J Woo and P. A. Monson, preprint (2002).
\bibitem{N1997} see the articles in {\it Spin Glasses and Random Fields}, 
A. P. Young, Ed. (World Scientific, Singapore, 1997).
\bibitem{KRTP1998}  E. Kierlik, M. L.  Rosinberg, G. Tarjus, and
E. Pitard, Mol. Phys. {\bf 95}, 341 (1998).
\bibitem{WSM2001} H-J. Woo, L. Sarkisov, and P. A. Monson,  Langmuir \textbf{17}, 7472 (2001).
\bibitem{DKRT2002} F. Detchevery, E. Kierlik, M. L. Rosinberg, and G. Tarjus (in preparation).
\bibitem{S1983} R. B. Stinchcombe, in {\it Phase transitions and
critical phenomena}, eds. C. Domb and J. L. Lebowitz (Academic Press,
London, 1983), vol.7, p. 151.
\bibitem{SP1992}The fact that a porous medium exerts  spatially
varying  fields  that break the hole-particle  symmetry
of  a lattice gas model (or the up-down symmetry of the equivalent
Ising spin model)  has also been noted by D. Stauffer and R. B. Pandey, J. Phys. A: Math. Gen.  {\bf 25}, L1079
(1992).
\bibitem{MSCCCB1991} It should be noted that  the hole-particle
symmetry of the model is broken even when the
probability distribution of the fields is symmetric, i.e., for
$\rho_m=1/2$. It is thus the  {\it combination} of dilution and
random fields that  is relevant. The asymmetric  random-field Ising model proposed by   A. Maritan {\it et al.},
Phys. Rev. Lett. {\bf 67}, 1821 (1991), does not include dilution and
cannot describe the influence of porosity on the fluid properties.
\bibitem{FGK1988} D. S. Fisher, G. M. Grinstein, and A. Khurana, 
Phys. Today {\bf 41} (12), 58 (1988).
\bibitem{IM1975} Y. Imry and S. K. Ma, Phys. Rev. Lett. {\bf 35}, 1399
(1975).
\bibitem{KMRT1997}  E. Kierlik, M. L.  Rosinberg, G. Tarjus, and
P. A. Monson, J. Chem. Phys. {\bf 95}, 264 (1997).
\bibitem{TAP1977}  D. J. Thouless, P. W. Anderson, and R. G. Palmer,
Philos. Mag. {\bf 35}, 593 (1977).
\bibitem{SM2002} L. Sarkisov and P. A. Monson, preprint (2002).
\bibitem{KRT2002} E. Kierlik, M. L.  Rosinberg, and G. Tarjus (in preparation)
\bibitem{LBL1983} D. D Ling, D. R. Bowman, and K. Levin,
Phys. Rev. B. {\bf 28}, 262 (1983).
\bibitem{LMP1995} D. Lancaster, E. Marinari, and G. Parisi, J. Phys. A: Math. Gen. {\bf 28}, 3959 (1995).
\bibitem{MPP2000} R. Mulet, A. Paganani, and G. Parisi, Phys. Rev. B. {\bf 63 }, 184438 (2001).
\bibitem{M1992} A. A. Middleton, Phys. Rev. Lett. {\bf 68}, 670
(1992); A. A. Middleton and D. S. Fisher, Phys. Rev. B {\bf 47}, 3530
(1993).
\bibitem{S1993}J. P. Sethna \emph{et al.}, Phys. Rev. Lett. \textbf{70}, 3347 (1993).
\bibitem{rem} This is still a small size compared to those
considered  in the studies of disordered magnets at $T=0$ (see, e.g.,
M. C. Kuntz  \emph{et al.}, Comp. Sci. Eng. {\bf 1}, 73 (1999)).
This of course is due to the fact that the local densities  $\rho_i$ are not restricted to the values $(0, 1)$, which forbids
the use of bits algorithms and requires large amounts of memory.
\bibitem{SA1994} D. Stauffer and A. Aharony, {\it Introduction to
Percolation Theory}, (Taylor and Francis, London, 1994).
\bibitem{WH2000} A. H. Wooters and R. B. Hallock,  J. Low
Temp. Phys. {\bf 121}, 549 (2000).
\bibitem{STC1999} R. Salazar, R. Toral, and A. Chakrabarti, J. Sol Gel
Sci. Technol. {\bf 15}, 175 (1999).
\bibitem{Rem} Note that the system does not stay  in
the same minimum along the whole equilibrium isotherm; in a
finite-size sample, this results in the small jumps
observed  in $\rho_f^{eq}(\mu)$. However, a  jump occurs at equilibrium when the
grand potentials associated with two local minima cross each other
(like in a usual first-order transition) whereas  a jump
occurs off-equilibrium when a local  minimum looses its stability
(like at a usual spinodal): for a finite-size sample, $\Omega^{ads}(\mu)$ and
$\Omega^{des}(\mu)$ are thus discontinuous curves whereas $\Omega^{eq}(\mu)$ is continuous.
\bibitem{DY1983} C. De Dominicis and A. P. Young, J. Phys. A: Math. Gen. {\bf 16}, 2063 (1983).
\bibitem{B1959} R. Brout, Phys. Rev.  {\bf 115}, 824 (1959). 
\bibitem{VP1994} E.   Vives and A. Planes, Phys. Rev. B.  \textbf{50},
3839 (1994); O.  Dahmen and J. P. Sethna, Phys. Rev.  \textbf{B 53},
14872 (1996), and   references   therein.

\end{references}
\end{document}